\documentclass[final]{aa}
\usepackage{my_epsf}
\newcommand{\bari}{{\scriptscriptstyle \hspace{-0.5em} \stackrel{-}{}}}
\newcommand{\beq}{\begin{equation}}
\newcommand{\eeq}{\end{equation}}
\newcommand{\beqarr}{\begin{eqnarray}}
\newcommand{\eeqarr}{\end{eqnarray}}
\newcommand{\dddot}[1]{\stackrel{\; \dots}{#1}}
\newcommand{\Ma}{M\!a^*}
\newcommand{\teq}{\!=\!}        

\def\gcc{\mathrm{g\,cm}^{-3}}

\def\etal{{\it et al.}\,}
\def\ie{{\it i.e.,}\,}
\def\eg{{\it e.g.,}\,}
\def\ms{\, {M_{\odot}}}

\begin{document}

\thesaurus{01(02.07.1, 02.08.1, 02.09,1, 08.14.1, 08.18.1, 08.19.4)}

\title{Simulations of non-axisymmetric rotational core collapse}

\author{M.~Rampp    \thanks{e-mail: {\tt mjr@mpa-garching.mpg.de}}
   \and E.~M\"uller \thanks{e-mail: {\tt emueller@mpa-garching.mpg.de}}
   \and M.~Ruffert  \thanks{e-mail: {\tt mruffert@ast.cam.ac.uk}}}

\offprints{M.~Rampp}

\institute{Max-Planck-Institut f\"ur Astrophysik,
           Karl-Schwarzschild-Str.~1, Postfach 1523,
           85740 Garching, Germany}

\date{November 11, 1997}

\maketitle

\begin{abstract}

We report on the first three-dimensional hydrodynamic simulations of
secular and dynamical non-axisymmetric instabilities in collapsing,
rapidly rotating stellar cores which extend well beyond core
bounce. The resulting gravitational radiation has been calculated
using the quadrupole approximation.

We find that secular instabilities do not occur during the simulated
time interval of several 10\,ms. Models which become dynamically
unstable during core collapse show a strong nonlinear growth of
non-axisymmetric instabilities. Both random and coherent large scale
initial perturbations eventually give rise to a dominant bar-like
deformation ($\exp(\pm im\phi$) with $m=2$). In spite of the
pronounced tri-axial deformation of certain parts of the core no
considerable enhancement of the gravitational radiation is found.
This is due to the fact that rapidly rotating cores re-expand after
core bounce on a dynamical time scale before non-axisymmetric
instabilities enter the nonlinear regime. Hence, when the core 
becomes tri-axial, it is no longer very compact.

\keywords{Gravitation -- Hydrodynamics -- Instabilities -- Stars:
          neutron -- Stars: rotation -- Supernovae: general}
\end{abstract}

\section{Introduction}

The rotation rate of a self-gravitating fluid body is usually
quantified by the parameter $\beta := E_{\rm rot}/|E_{\rm
pot}|$, where $E_{\rm rot}$ is the rotational energy and $E_{\rm pot}$ the potential energy of the body (\eg Tassoul
1978).  It is well known from linear stability analysis that global
non-axisymmetric instabilities can develop in a rotating body when
$\beta$ is sufficiently large (\eg Chandrasekhar 1969; Tassoul 1978).
Rotational instability arises from non-radial azimuthal (or toroidal)
modes $\exp(\pm im\phi)$, where $\phi$ is the azimuthal coordinate and
where $m$ is characterizing the mode of perturbation. The mode with
$m=2$ is known as the ``bar'' mode.  Driven by hydrodynamics and
gravity, a rotating fluid body becomes dynamically unstable to
non-axisymmetric perturbations on a time scale of approximately one
rotation period when $\beta > \beta_{\rm dyn}$. Secular instabilities
develop due to dissipative processes such as gravitational radiation
reaction (GRR; Chandrasekhar 1970), viscosity (Roberts \& Stewartson
1963), or a combination of both (Lindblom \& Detweiler 1977),
and grow on a time scale of many rotation periods (Schutz 1989)
when $\beta_{\rm sec} < \beta < \beta_{\rm dyn}$.

The values of the critical rotation parameters $\beta_{\rm sec}$ and
$\beta_{\rm dyn}$ depend on the density stratification and the angular
momentum distribution of the rotating body.  For MacLaurin spheroids,
\ie for incompressible, rigidly rotating equilibrium configurations,
secular and dynamical bar instabilities set in for $\beta_{\rm sec} =
0.1375$ and $\beta_{\rm dyn} = 0.2738$, respectively (\eg
Chandrasekhar 1969; Tassoul 1978).  In differentially rotating
polytropes ($p \propto \varrho^{1+1/n}$) both the angular momentum
distribution and to a lesser extent the polytropic index $n$ affect
the value of $\beta_{\rm sec}$ at which the secular bar mode sets in
(Imamura \etal 1995).  For $n=3/2$ polytropes and $m=2$ perturbations
Pickett \etal (1996) find that the classical dynamical stability limit
$\beta_{\rm dyn} \approx 0.27$ holds, if the angular momentum
distribution is similar to those of MacLaurin spheroids.  However,
considerably lower values of $\beta_{\rm dyn}$ are found when the
latter gives rise to extended Keplerian-disk-like equilibria.

It is unclear to what extent these results obtained analytically for
Mac\-Laurin spheroids and numerically for compressible configurations
are applicable to collapsing cores, since they are derived for
stationary {\em equilibrium} models.  It has been shown, however, that
during axisymmetric rotational core collapse conservation of angular
momentum can lead to very rapidly rotating configurations, whose
rotation rates exceed $\beta_{\rm sec}$ and even $\beta_{\rm dyn}$
(Tohline 1984; Eriguchi \& M\"uller 1985; M\"onch\-meyer \etal 1991;
Zwerger \& M\"uller 1997).  But whether such super-critical rotation
in collapsing cores indeed leads to the growth of non-axisymmetric
instabilities has not yet been demonstrated in hydrodynamic
simulations.

Besides of possibly being relevant for the collapse dynamics of
rapidly rotating stellar cores, tri-axial instabilities have also been
envisaged to boost the gravitational wave signal from rotational core
collapse (for a review, see \eg \- Piran 1990; Thorne 1995). This
would be of great importance for the four long-baseline laser
interferometric gravitational wave detectors (GEO600, LIGO, TAMA,
VIRGO) which will become operational within the next few years. They
are designed to achieve sensitivities (for single bursts) down to
gravitational wave amplitudes of $|h| \sim 10^{-22}$ (see \eg
Abramovici \etal 1992).  In order to obtain a few events per year, one
has to detect all core collapse supernovae out to the Virgo cluster of
galaxies (distance $\simeq 10$~Mpc).  
However, simulations of axisymmetric
rotational core collapse predict gravitational wave amplitudes of at
most $|h| \sim 10^{-23}$ at that distance (M\"uller 1982; Finn \&
Evans 1990; M\"onch\-meyer \etal 1991; Yamada \& Sato 1995; Zwerger \&
M\"uller 1997). Significantly larger gravitational wave signals could
be produced if non-axisymmetric instabilities due to rotation act in
the collapsing core.  According to this idea, a rapidly spinning core
will experience a centrifugal hang-up and will be transformed into a
bar-like configuration that spins end-over-end like an American
football, if its rotation rate exceeds the critical rotation rate(s).
One has further speculated, that the core might even break up into two
or more massive pieces, if $\beta > \beta_{\rm dyn}$ (\eg Bonnell \&
Pringle 1995).  It has been suggested that the resulting gravitational
radiation {\it could} be almost as strong as that from coalescing
neutron star binaries (Thorne 1995).  The actual strength of the
gravitational wave signal will sensitively depend on (i) the radius at
which the centrifugal hang-up occurs and (ii) what fraction of the
angular momentum of the non-axisymmetric core goes into gravitational
waves, and what fraction into hydrodynamic waves.  These sound and
shock waves are produced as the bar or lumps, acting like a
twirling-stick, plow through the surrounding mass layers.

Recently, both semi-analytic and numerical methods have been used to
compute the gravitational radiation produced by non-axisymmetric
instabilities in rapidly rotating stars. Lai \& Shapiro (1995) have
studied secular instabilities in rapidly rotating neutron stars and
have computed the resulting gravitational radiation using linearized
dynamical equations and a compressible ellipsoid model. Houser \etal
(1994), Houser \& Centrella (1996) and Smith \etal (1996) have used
three-dimensional hydrodynamic codes to simulate the nonlinear growth
of the dynamical tri-axial instability in rapidly ($\beta = 0.3$)
rotating polytropes ($n=3/2, n=1, n=1/2$). Scaling their results to
neutron star dimensions (\ie a polytrope with mass $M \sim 1.4\,\ms$
and radius $R \sim 10$\,km) they have also calculated the
gravitational radiation from the non-axisymmetric instability. They
obtained a maximum dimensionless gravitational wave amplitude $|h|
\sim 2\,10^{-22}$ for a source at a distance of 10\,Mpc, the energy
lost to gravitational radiation being $\Delta E \sim 10^{-3}\,M c^2$.

Concerning investigations, like those just discussed, it is important
to note that they can be used to predict the gravitational radiation
from rapidly rotating, stationary neutron stars, which might (or might
not!) form as a consequence of rotational core collapse.  They are not
appropriate, however, for predicting the gravitational wave signature
of the {\em collapse} of rapidly rotating stellar cores.  
We stress this difference here, because it is often overlooked.

Up to now three-dimensional hydrodynamic simulations of rotational
core collapse, which can follow the nonlinear growth of
non-axisymmetric instabilities {\em during} collapse, have only been
performed by Bonazzola \& Marck (1993, 1994), and by Marck \&
Bonazzola (1992). For their simulations they used a pseudo-spectral
hydrodynamic \- code and a polytropic equation of state.  They
computed the evolution of several initial models and found that the
gravitational wave amplitude is within a factor of two of that of 2D
simulations for the same initial deformation of the core (Bonazzola \&
Marck 1994).  However, their simulations were restricted to the
pre-bounce phase of the collapse, and thus are less relevant.

In the following we report on the first three-dimensio\-nal hydrodynamic
simulations of non-axisymmetric instabilities in collapsing, rapidly
rotating stellar cores which extend well beyond core bounce. In
addition to the dynamics, we have also computed the gravitational
radiation emitted during the evolution. Our study is a continuation
and an extension of the recent work of Zwerger (1995) and Zwerger \&
M\"uller (1997), who have performed a comprehensive parameter study of
axisymmetric rotational core collapse. The initial models, the
equation of state and the hydrodynamic method are adopted from their
study. From their set of 78 models we have only considered those which
exceed the critical rotation rate(s) during core collapse.

The paper is organized as follows: In section\,2 we pre\-sent the
computational procedure used to follow the hydrodynamic evolution of
the core and to compute the gravitational wave signal. Subsequently,
in section\,3, we first discuss the results of a two-dimensional
simulation, which serves as a reference point for the
three-dimensional runs. We then pre\-sent various aspects of the
evolution of the non-axisymmetric models and discuss the gravitational
wave signature of the models. Finally, in section\,4, we summarize our
results, and conclude with a discussion of the shortcomings of our
approach.

\section{Computational procedure}

\subsection{Hydrodynamic method and equation of state}

The simulations were performed with the Newtonian mul\-ti-dimensional
finite-volume hydrodynamic code PROMETHEUS developed by Bruce Fryxell
and Ewald M\"uller (Fryxell \etal 1989). PROMETHEUS is a direct
Eulerian implementation of the Piecewise Parabolic Method (PPM) of
Colella \& Woodward (1984). For the three-dimensional simulations we
used a variant of PROMETHEUS due to Ruffert (1992), which utilizes
multiple-nested refined equidistant Cartesian grids to enhance the
spatial resolution. We do not include any general relativistic effects
for the fluid and have neglected effects due to neutrino transport.

Matter in the core is approximated by a perfect fluid using a
simplified analytic equation of state (Janka \etal 1993; see also
Zwerger \& M\"uller 1997 for details of the implementation).  The
pressure is a function of the density $\varrho$ and energy density $u$
of the core matter and consists of two parts:

\beq
\label{EOS1}
P(\varrho,u) = P_{\rm p} + P_{\rm th} \,.
\eeq

The polytropic part

\beq
\label{P_p}
P_{\rm p} = K\cdot\varrho^{\Gamma_{\rm p}} \,,
\eeq

which depends only on density, describes the pressure contribution of
degenerate relativistic electrons or that due to the repulsive nuclear
forces at high densities.  The thermal part $P_{\rm th}$ mimics the
thermal pressure of shock-heated matter. It is computed from the
corresponding internal energy density $u_{\rm th}$ by

\beq
\label{P_th}
P_{\rm th} = (\Gamma_{\rm th}-1) u_{\rm th},
\qquad \mbox{with}\qquad
\Gamma_{\rm th} = 1.5 \,.
\eeq

The thermal energy density $u_{\rm th}$ in turn is given by the total
energy density $u$ through the relation

\beq
u = u_{\rm p}+u_{\rm th} \,,
\eeq

where $u_{\rm p}$ is the energy density of the degenerate electron
gas.  To account for the repulsive part of the nuclear forces the
``polytropic'' adiabatic exponent $\Gamma_{\rm p}$ (see
Eq.\,(\ref{P_p})) is increased from a value $\Gamma_1$ close to $4/3$
(resembling degenerate, relativistic electrons) to a value
$\Gamma_2=2.5$, if the density of a fluid element exceeds nuclear
matter density which is set to $\varrho_{\rm nuc} = 2.0\,10^{14}\gcc$.

\subsection{Gravitational wave emission}\label{GW}

In the transverse-traceless (TT) gauge a metric perturbation $\tens h$
can be decomposed as (\eg Misner \etal 1973, chap.~35)

\beq\label{h_TT}
\tens{h}^{\rm TT}=h_+\tens{e}_+ +h_{\times}\tens{e}_{\times}
\eeq

with the unit linear polarization tensors defined, in spherical
coordinates $(r, \theta, \phi)$, as

\beqarr
\tens{e}_{+}      &=& \vec{e}_{\theta} \otimes
                      \vec{e}_{\theta}-\vec{e}_{\phi} \otimes
                      \vec{e}_{\phi}
\nonumber \\
\tens{e}_{\times} &=& \vec{e}_{\theta} \otimes
                      \vec{e}_{\phi}+\vec{e}_{\phi} \otimes
                      \vec{e}_{\theta}  \,,
\eeqarr

where $\vec{e}_{\theta}$ and $\vec{e}_{\phi}$ are the unit vectors in
the corresponding coordinate directions and $\otimes$ is the tensor
product.  The two independent polarizations $h_+$ and $h_\times$ are
calculated from the fluid variables in Post-Newtonian approximation,
keeping only the mass-quadrupole in the multipole expansion of the
field.  Let us consider an observer located at coordinates $(r,\theta,
\phi)$ in a spherical coordinate system whose origin coincides with
the center of mass of the core. Then the two independent polarizations
are given by (\eg Zhuge \etal 1994)

\begin{eqnarray}
h_{+} = & {\displaystyle \frac{ G}{c^4}\frac{1}{r}} &
           \Big(\: ( \ddot I {\bari} _{xx}\cos^2\phi
         + \ddot I {\bari} _{yy}\sin^2\phi
         + \ddot I {\bari} _{xy}\sin2\phi)\cos^2\theta
\nonumber \\
     & & + \ddot I {\bari}_{zz}\sin^2\theta
         - (\ddot I {\bari}_{xz}\cos\phi
           +\ddot I {\bari}_{yz}\sin\phi)\sin 2\theta
\nonumber \\
     & & - \ddot I {\bari}_{xx}\sin^2\phi
         - \ddot I {\bari}_{yy}\cos^2\phi
         + \ddot I {\bari}_{xy}\sin 2\phi\:\Big)
\label{hplus}
\\
     & & \nonumber
\\
h_{\times} = & {\displaystyle \frac{G}{c^4}\frac{2}{ r}} &
                \Big(\:\frac{ 1}{ 2}( \ddot I {\bari} _{yy}
              - \ddot I {\bari} _{xx})\sin 2\phi\cos\theta
              + \ddot I {\bari} _{xy}\cos 2\phi\cos\theta
\nonumber \\
          & & +(\ddot I {\bari} _{xz}\sin\phi
              - \ddot I {\bari} _{yz}\cos\phi)\sin\theta\:\Big) \,.
\label{hcross}
\end{eqnarray}

Here the quantities $\ddot I {\bari}_{ij}$ are the second time
derivatives of the Cartesian components of the reduced quadrupole
moment tensor $\tens{I} {\hspace{0.12em}\bari}$. They are defined according to

\beq\label{d2I}
\ddot I {\bari}_{ij}:=\frac{{\rm d}^2}{{\rm d}t^2}
                      \int\varrho\left(\vec{x},t-r/c\right)
                      \cdot \left(x_i
                           x_j-\frac{1}{3}\delta_{ij}x_kx^k\right)
                      \,{\rm d}^3\vec{x} \,,
\eeq

where we have used the Einstein sum convention (\ie\- summation over
repeated indices) and where $\delta_{ij}$ is the Kronecker
symbol. Note that the origins of the spherical coordinate system (used
to specify the observer's position) and the Cartesian coordinate
system (used to describe the matter distribution of the source)
coincide. The coordinate systems are oriented relative to each other
such that $\theta=0$ corresponds to the positive $z$-direction.

We assume that the source possesses equatorial symmetry (\ie it is
symmetric with respect to the transformation $z$\,$\to$\,$-z$), and
hence $I {\bari}_{xz}=I {\bari} _{yz}\equiv 0$.  Inspection of the
remaining terms in Eqns.\,(\ref{hplus}) and (\ref{hcross}) then yields
$\theta\in\{0,\pi/2\}$ and $\theta=0$ as necessary conditions for
extrema of the waveforms $h_{+}$ and $h_{\times}$ (considered as
functions of the observer's position), respectively.  Thus, we
calculate $h_{\times}$ along the positive $z$-axis and $h_{+}$ in the
equatorial plane (at an azimuthal position $\phi=0$) to yield the
maximum stresses\footnote{For all models discussed below, the value of
$h_{+}$ at $\theta=0$ turned out to be negligibly small.}.  We have
checked for all models that we do not miss possible additional (short
term) maxima of $h_+$ that could be present for an observer at some
azimuthal position $\phi\ne 0$ in the equatorial plane but could be
overlooked by an observer located at $\phi=0$.

The quadrupole part of the total gravitational wave energy is given by

\beq
\label{E_GW}
E_{\rm GW} = \frac{1}{5}\frac{G}{c^5}
             \int^\infty_{-\infty}\dddot {I \bari} _{ij}
                                  \dddot {I \bari} _{ij}\,{\rm d}t \,.
\eeq

Instead of numerically approximating Eq.\,(\ref{d2I}) with standard
finite differences, it is superior (\eg Finn \& Evans 1990;
M\"onch\-meyer 1993) to use an equivalent expression derived
independently by Nakamura \& Oohara (1989) and by Blanchet \etal
(1990):

\beq
\label{d2I_num}
\ddot I {\bari}_{ij} = {\rm STF}\left\{2
               \noindent
        \int\varrho \cdot (v^iv^j-x^i\partial_j\Phi)
                       \,{\rm d}^3\vec{x} \right\} \,,
\eeq

where the symmetric and trace free part of a doubly indexed quantity
$A_{ij}$ is defined by

\beq
{\rm STF}\{A^{ij}\}:=  \frac{1}{2}A^{ij}
                      +\frac{1}{2}A^{ji}
                      -\frac{1}{3}\delta^{ij}A^{ll} \,.
\eeq

The spatial derivatives $\partial_j\Phi$ of the gravitational
potential appearing in Eq.\,(\ref{d2I_num}) are approximated by
centered finite differences.

For two reasons no back-reaction of the gravitational wave emission on
the fluid (GRR) has been implemented. Firstly, the peak luminosity of
gravitational radiation that can be expected from rotational core
collapse is at most of the order of $10^{50}$\,erg\,s$^{-1}$ (M\"uller 1982;
Finn \& Evans 1990; M\"onch\-meyer \etal 1991; Zwerger \& M\"uller
1997). When comparing this luminosity with the typical energies
(kinetic, rotational, potential, internal) $E \simeq 10^{52}$\,erg of
the core one finds a GRR time scale of the order of 100\,s, which is
at least by a factor of $10^4$ larger than the time interval over which
we can calculate the core's evolution.  Secondly, total energy
conservation is violated by an amount $\Delta E \simeq 10^{49}$\,erg
in the best resolved 2D calculation and by an amount $\Delta E \simeq
10^{50}$\,erg in the 3D runs. This violation means an acceptable
1\%-error as far as the dynamics is concerned. However, it dominates
the change of energy radiated due to GRR ($E_{\rm GW}\la
10^{47}$\,erg) in the time interval of our simulations by about a
factor of $10^3$.

\subsection{Two-dimensional simulations}

Closely following Zwerger \& M\"uller (1997), the axisymmetric
(reference) simulations were performed in spherical coordinates assuming
equatorial symmetry.  Thus, the angular grid covered the range $\theta
\in [0,\pi/2]$.  The number of equidistant angular zones was $n_\theta =
18$, 45 or 90, which corresponds to an angular resolution of $5^\circ$,
$2^\circ$ or $1^\circ$, respectively.  The moving radial grid consisted
of 360 zones.  The grid was moved in such a way that the inner core (\ie
the subsonic inner part of the core) was always resolved by 180 radial
zones.  The radial zoning varied from about 0.5\,km in the unshocked
inner core to about 50\,km in the outermost parts of the outer core (for
more details see Zwerger \& M\"uller 1997).

As in Zwerger \& M\"uller (1997), the gravitational wave amplitude was
computed by expanding the gravitational field into ``pure-spin tensor
harmonics'' (Thorne 1980). The only non-vanishing quadrupole
contribution is then given by

\beq
\label{AE2_20}
A^{{\rm E}2}_{20} = \frac{G}{c^4} \frac{16\pi^{3/2}}
                    {\sqrt{15}} \, \frac{{\rm d}^2}{{\rm d}t^2}
                    \int_0^\infty\int_{-1}^1\varrho
                    \left(\frac{3}{2}z^2-\frac{1}{2}\right)r^4
                    \,{\rm d}z{\rm d}r  \,,
\eeq

which is related to the (dimensionless) metric perturbations of
Eq.\,(\ref{h_TT}) by

\beq
h_+ = \frac{1}{8}\sqrt{\frac{15}{\pi}}\sin^2\theta\,\,
      \frac{A^{{\rm E}2}_{20}}{r} \,,
\qquad
h_{\times}\equiv 0
\eeq

Using the counterpart of Eq.\,(\ref{d2I_num}) in spherical coordinates
(\eg Zwerger \& M\"uller 1997, Eq.\,20), the second time derivative as
well as the $r^2$-weighting of mass elements, both numerically
troublesome (\eg Finn \& Evans 1990), can be eliminated.

In the quadrupole approximation the total gravitational wave energy is
given by

\beq
E_{\rm GW} = \frac{c^3}{G}\frac{1}{32\pi}\int^\infty_{-\infty}
             \left(\frac{{\rm d}A^{{\rm E}2}_{20}}{{\rm d}t}\right)^2
             \,{\rm d}t  \,.
\eeq

Higher-order terms in the multipole expansion of the gravitational
wave field are negligible both for the wave amplitudes and the amount
of radiated energy in the axisymmetric case (\eg M\"onch\-meyer \etal
1991, Zwerger 1995).

\begin{figure*}[!t]
 \begin{tabular}{cc}
\put(0.9,0.3){{\LARGE\bf a}}
\epsfxsize=8.8cm \epsfclipon \epsffile{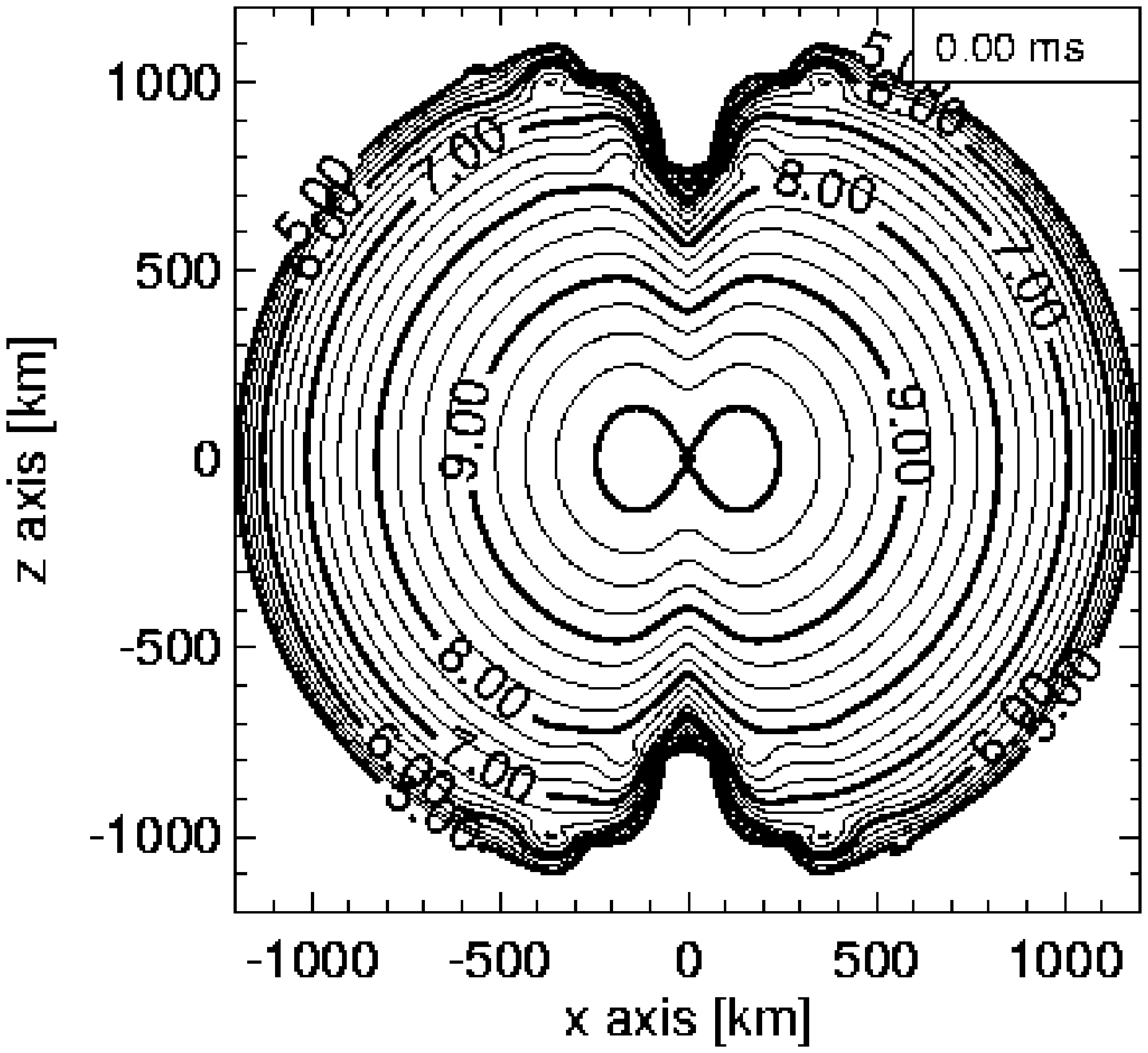} &
\put(0.9,0.3){{\LARGE\bf b}}
\epsfxsize=8.8cm \epsfclipon \epsffile{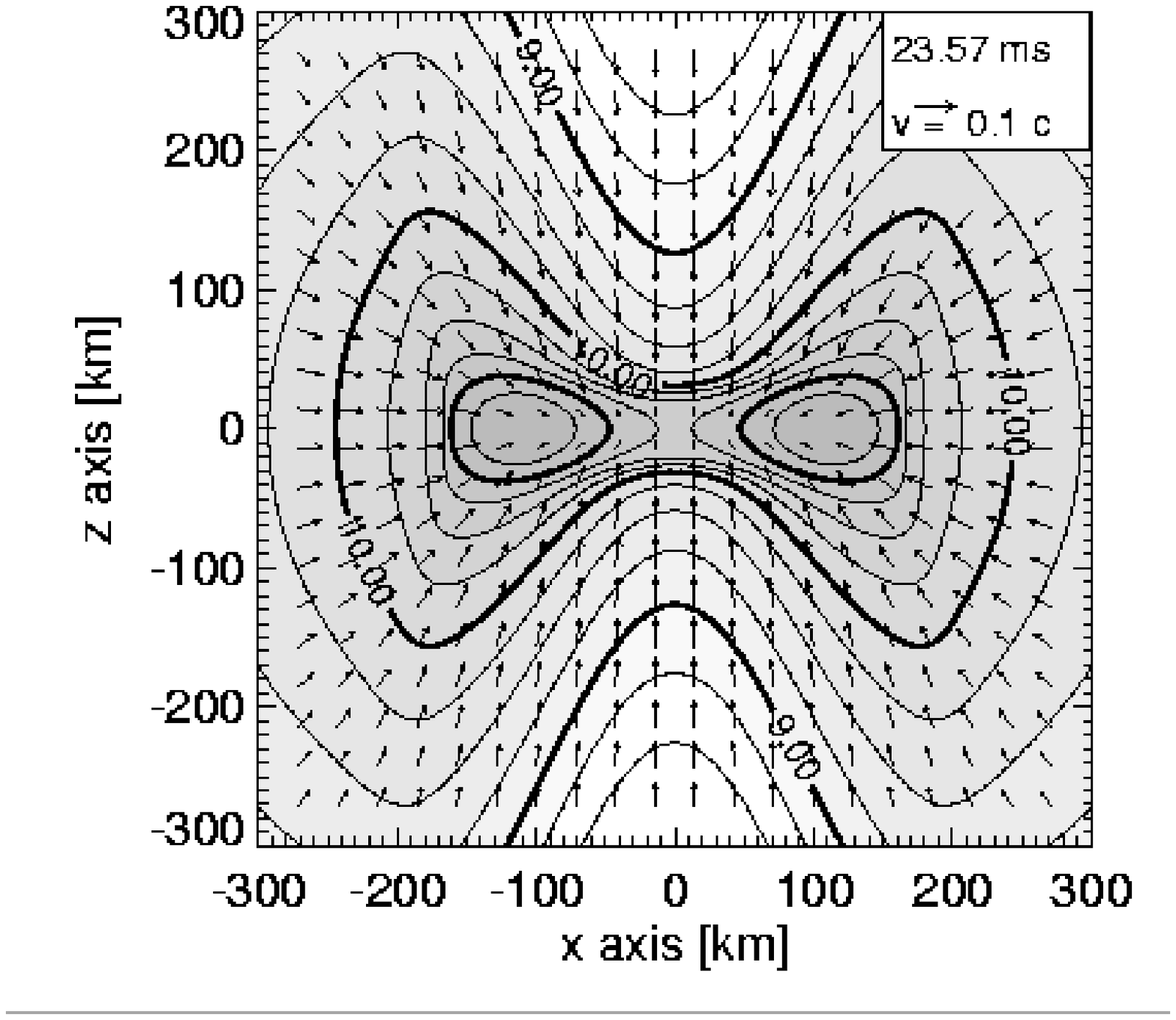} \\ [-2ex]
\put(0.9,0.3){{\LARGE\bf c}}
\epsfxsize=8.8cm \epsfclipon \epsffile{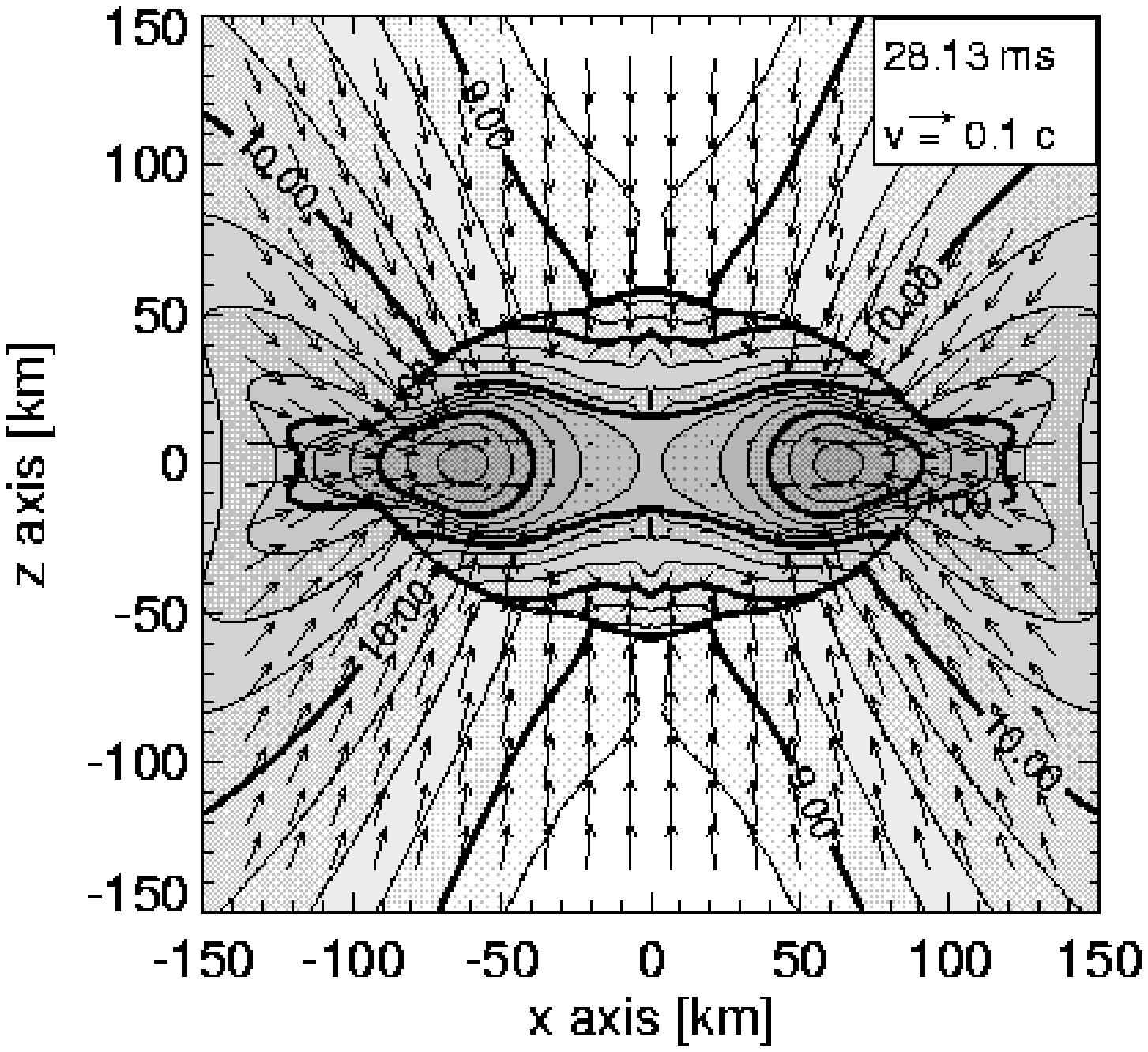} &
\put(0.9,0.3){{\LARGE\bf d}}
\epsfxsize=8.8cm \epsfclipon \epsffile{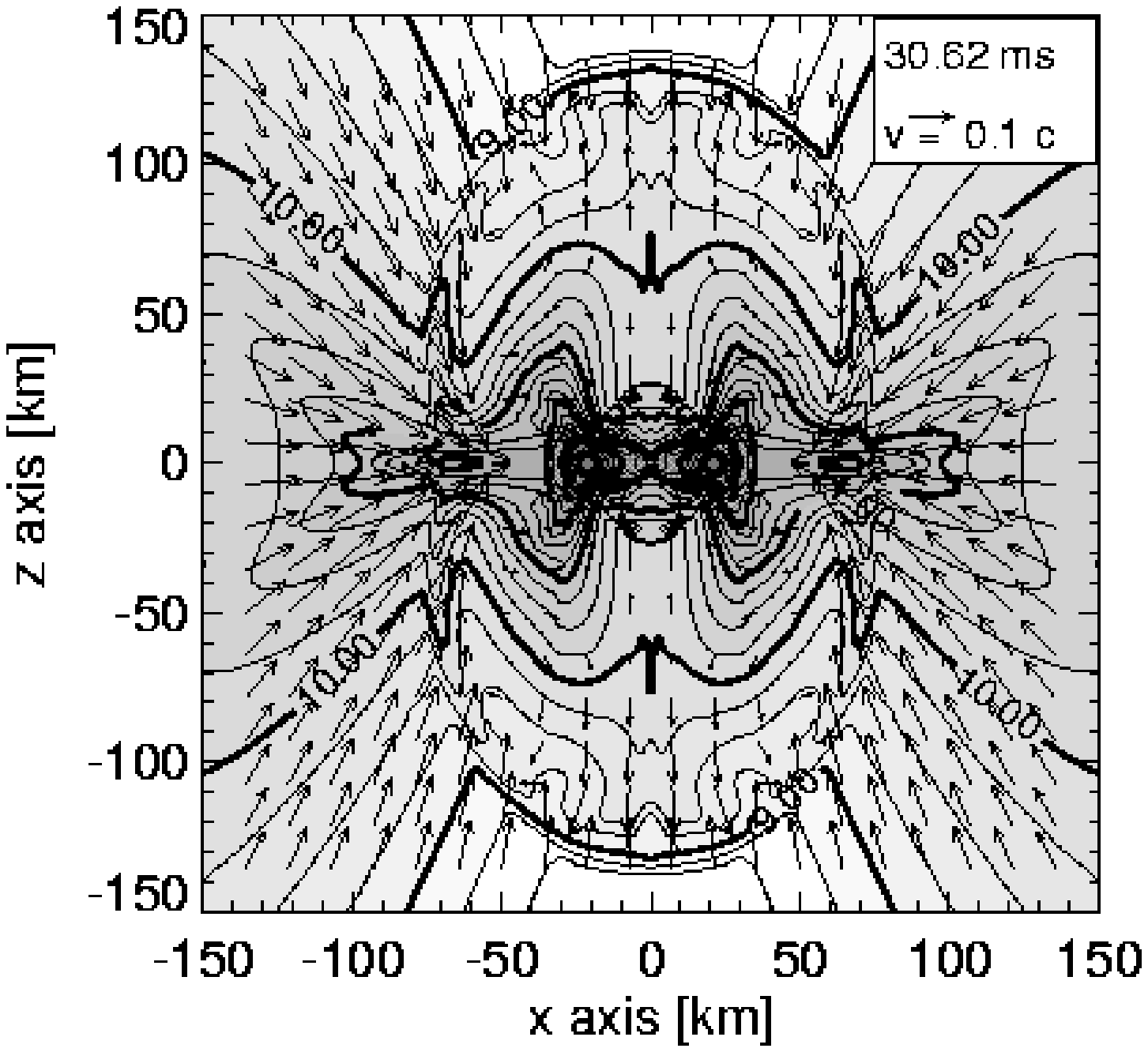} \\ [-2ex]
 \end{tabular}
\caption[]{Snapshots of the density distribution (in units of
[$\gcc$]) and the velocity field in a meridional plane for the
axisymmetric model A4B5G5. The $z$-axis coincides with the axis of
rotation (and symmetry). The contours are logarithmically spaced with
intervals of 0.25 dex, they are shaded with darker grey values for
higher density regions and labeled with their respective values. The
time of the snapshot and the velocity scale are given in the top right
corner of each panel.}
\label{fig:2D_cont}
\end{figure*}

\subsection{Three-dimensional simulations}

The three-dimensional calculations have been performed on
multiple-nested refined equidistant Cartesian grids (see Ruffert
(1992) for details of the method).  As in the 2D calculations we
assume the equatorial plane $z=0$ to be a plane of symmetry.  Six
centered grids were used each of which was refined by a factor of two.
In direction of the rotation ($z$-) axis each grid had 32\,zones,
while the perpendicular grid planes consisted of $64\times 64$ zones.
The coarsest grid had a spatial extent of 2560\,km covering the entire
core with cubes of edge size 40\,km.  The finest grid covered the
innermost 80\,km with a linear resolution of about 1\,km. Thus, the
resulting spatial resolution was comparable to that of the 2D
simulations.

For the three-dimensional simulations we considered axisymmetric models
of Zwerger \& M\"uller (1997), which are stable against non-axisymmetric
perturbations in the early stages of collapse (\ie $\beta(t) \la 0.1$),
but later exceed the critical rotation rate(s).  
Test calculations
showed that non-axisymmetric instabilities did not grow in perturbed
axisymmetric models with $\beta \la 0.1$ on time scales of a few
10\,ms.  
Using trilinear interpolation an axisymmetric model was mapped
onto the three-dimensional grid when its rotation parameter reached a
value of $\approx 0.1$. Typically, this occured a few milliseconds
before core bounce.

We then imposed certain non-axisymmetric perturbations and followed the
subsequent dynamical evolution of the core in three spatial dimensions.
Due to the cubic cells of our computational grid, the mapping procedure
itself introduces deviations from axisymmetry that are symmetric with
respect to the coordinate transformations

\beq
\phi\to\phi+\pi/2\qquad \mbox{and} \qquad \phi\to\phi+\pi \,.
\eeq

This implies that only azimuthal modes $\exp(im\phi)$ with even values
of $m$ are contained in the ``grid-mapping noise''.  The amplitudes
are of the order of a few percent measured as relative deviations
$\delta\psi := (\psi-\langle\psi\rangle_\phi) \,/\,
\langle\psi\rangle_\phi$ of a scalar field $\psi(r,\theta,\phi)$ from
its azimuthal mean $\langle\psi\rangle_\phi := (2\pi)^{-1} \,
\int_0^{2\pi} \psi(r,\theta,\phi) \, {\rm d}\phi$.  This defines a
lower limit for the size of the perturbation amplitudes, which we can
impose explicitly onto the axisymmetric models.  We point out that
density perturbations $\delta \varrho \simeq 1\%$ are about two orders
of magnitude larger than those typically used in numerical stability
analysis (\eg Houser \etal 1994; Aksenov 1996; Smith \etal 1996).

We used explicit random perturbations of the density with an amplitude
of 10\%, \ie the density distribution of the mapped model was modified
according to

\beq
\varrho(r,\theta,\phi)\to \varrho(r,\theta,\phi)
\,\cdot\, \big[1+0.1\,f(r,\theta,\phi)\big]  \,,
\eeq

where $f$ denotes a random function distributed uniformly in the
interval $[-1,1]$.  All azimuthal modes that can be resolved on a given
grid are therefore present in the initial data, the $m=4$ and $m=2$
``grid-modes'' mentioned above presumably being dominant.  The
corresponding model is refered to in the following as model MD1.

In a second model MD2, we imposed in addition a large scale $m=3$
azimuthal perturbation with an amplitude of 5\%, \ie

\beq
\varrho(r,\theta,\phi)\to \varrho(r,\theta,\phi)
\,\cdot\, \big[1+0.05\,\sin(0.6+3\phi)\big]   \,.
\eeq

Note that for convenience of notation we have given the form of the
perturbation in spherical coordinates, although it is introduced on a
Cartesian grid.

\section{Results}

Zwerger \& M\"uller (1997) have studied the overall dynamics of
axisymmetric rotational core collapse and calculated the resulting
gravitational wave signal.  In particular, they have investigated how
the dynamics and the gravitational radiation depend on the (unknown)
initial amount and distribution of the core's angular momentum, and on
the equation of state.  Among the 78 models computed by them, Zwerger \&
M\"uller (1997) found only one model (A4B5G5), whose rotation rate
parameter considerably exceeds $\beta_{\rm dyn}$ near core bounce.
However, $\beta$ remains larger than $\beta_{\rm dyn}$ for only roughly
one millisecond, because the core rapidly re-expands after bounce and
hence slows down.  In addition, Zwerger \& M\"uller (1997) found three
more models (A3B5G4, A3B5G5 and A4B5G4) that fulfilled $\beta>\beta_{\rm
sec}$ for several 10\,ms (see Fig.\,4 of Zwerger \& M\"uller 1997).

The two ``classical'' dissipation mechanisms that can in principle
drive secular non-axisymmetric instabilities, namely viscosity of core
matter (Roberts \& Stewartson 1963) and gravitational radiation
reaction (GRR; Chandrasekhar 1970), are negligible during the time
scales we have simulated evolution of the core (a few 10\,ms).
Nevertheless, some inner fraction of the collapsing core could still
become unstable, because of acoustic coupling with or due to advection
of matter into the outer core (see \eg Schutz 1983 and references
therein).  Therefore, we have experimented with two of the three
models exceeding $\beta_{\rm sec}$ (but not $\beta_{\rm dyn}$) for
several 10\,ms by imposing non-axisymmetric perturbations on different
hydrodynamic \- quantities ($\varrho$, $P$, $\vec{v}$) of different
spatial character ($m=2$, random) and amplitude (5\%, 10\%) at
different epochs of the evolution.  On a time scale of several 10\,ms
we have found neither indications for secular instabilities nor a
significant enhancement of the gravitational wave signal (compared
with the corresponding axisymmetric model).

Thus, the only model remaining of the set of models of Zwerger \&
M\"uller (1997), which is a promising candidate for the growth of
tri-axial instabilities during its gravitational collapse, is model
A4B5G5 (see above).  Its properties are described in the next
subsection.  Some global quantities of the model have already been
published by Zwerger \& M\"uller (1997).

\begin{figure}[!t]
 \epsfxsize=8.8cm \epsfclipon \epsffile{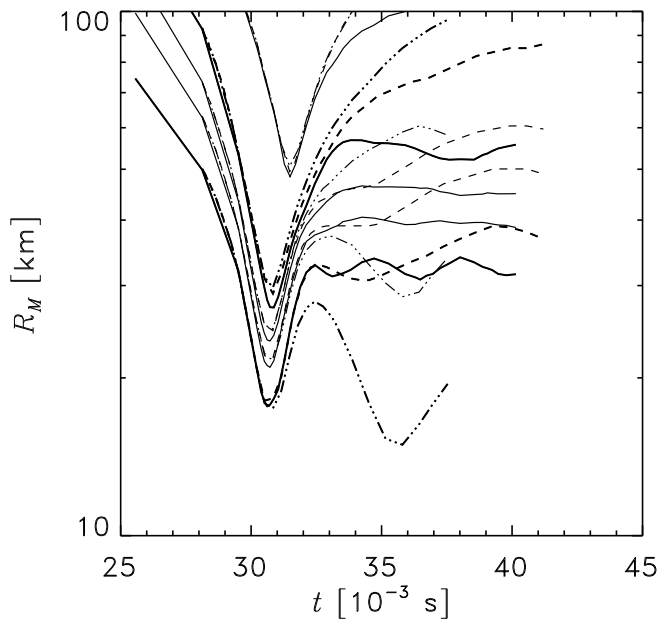}
\caption[]{Radii of selected mass shells (for their definition see
Eq.\,(\ref{m_r}) in the text) as a function of time for two three
dimensional models MD1 (dashed lines) and MD2 (dashed-dotted lines), and
the axisymmetric model A4B5G5 (solid lines).  The mass inside the shown
mass shells ranges from $0.1\ms$ to $0.9\ms$ in intervals of $0.2\ms$.
Lines corresponding to $0.1\ms$ and $0.7\ms$ are drawn
in bold.}
\label{fig:m_r}
\end{figure}

\subsection{The axisymmetric model A4B5G5}

All pre-collapse models of Zwerger \& M\"uller (1997) are axisymmetric
$n = 3$ polytropes ($p \propto \varrho^{1+1/n}$) in rotational
equilibrium whose angular velocity $\Omega$ only depends on the distance
$\varpi$ from the axis of rotation according to the rotation law

\beq
\label{rot}
\Omega(\varpi) \propto \left(1+(\frac{\varpi}{A})^2\right)^{-1} \,.
\eeq

Collapse is initiated by reducing the adiabatic index $\Gamma$ from the
value 4/3, which is consistent with the structure of a $n=3$ polytrope,
to a somewhat smaller value $\Gamma_1$ (see Eq.\,(\ref{P_p})).

The initial model A4B5G5\footnote{If not stated otherwise, quoted
numbers always refer to the 2D simulation with the best angular
resolution, \ie the one performed with $n_\theta=90$ (see
section\,2.3).} is the most extreme one in the large parameter set
considered by Zwerger \& M\"uller (1997).  It is the {\em fastest}
($\beta_{\rm i}=0.04$) and {\em most differentially} ($A=10^7$\,cm;
see Eq.\,(\ref{rot})) rotating model being evolved with the {\em
softest equation of state} ($\Gamma_1=1.28$).  According to the
criterion derived by Ledoux (1945; Eq.\,77) for rotating stars, a
rotation rate of $\beta = 0.04$ requires $\Gamma_1 > 1.3$ for the star
to be stable against the fundamental radial mode, \ie collapse.  The
resulting model has a mass of $1.66 M_\odot$, a total angular momentum
of $L_z = 3.85\,10^{49}$\,erg\,s and an initial equatorial radius of
1280\,km.

Fig.\,\ref{fig:2D_cont}a shows that model A4B5G5 has a torus-like
density stratification with an off-center density maximum, which is
located in the equatorial plane 200\,km away from the rotation axis.
During collapse a kind of ``infall channel'' forms along the rotation
axis, a structure which has already been found in some models of
M\"onch\-meyer \etal (1991).  There matter falls almost at the speed
of free-fall towards the center.  With decreasing latitude (\ie
in\-creasing polar angle $\theta$) matter is increasingly decelerated
by centrifugal forces (Fig.\,\ref{fig:2D_cont}b).  At $t \approx
28$\,ms, which is about 3\,ms before core bounce, an oblate shock is
visible (Fig.\,\ref{fig:2D_cont}c).  It begins already to form at the
bottom of the infall channel 7\,ms before core bounce
(Fig.\,\ref{fig:2D_cont}b).  Pressure gradients decreasing with
increasing polar angle $\theta$ lead to an asymmetric propagation of
the shock its speed being larger in polar than in equatorial regions.
The result is a ``prolate shock in an oblate star'' (Finn \& Evans
1990).


\begin{figure}[t]
 \epsfxsize=8.8cm \epsfclipon \epsffile{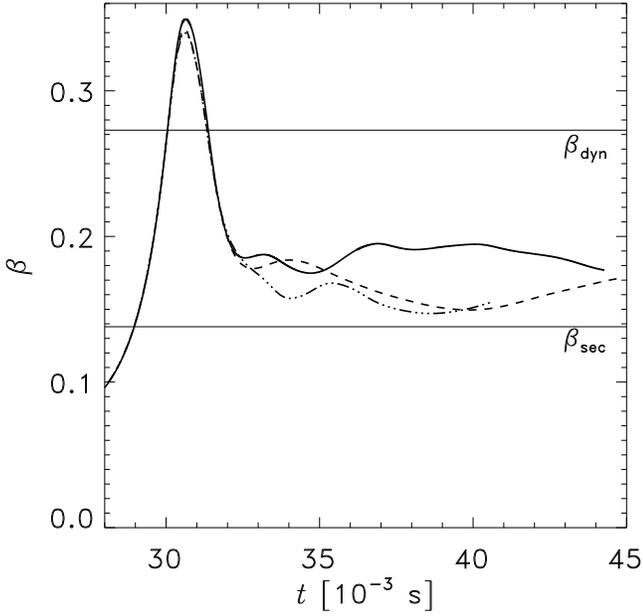}
\caption[]{Rotation rate parameter $\beta$ as a function of time for
the three-dimensional models MD1 (dashed) and MD2 (dashed-dotted).
For comparison, the evolution of $\beta$ of the two-dimensional model is
shown, too (solid). The solid horizontal lines mark the critical values
$\beta_{\rm sec}$ (lower line) and $\beta_{\rm dyn}$ (upper line),
where MacLaurin spheroids become secularly and dynamically unstable
to non-axisymmetric bar-like perturbations.}
\label{fig:3Dbeta}
\end{figure}
A second shock, which forms at the
edge of the torus shortly before the density in the core reaches its
maximum value inside the torus, propagates behind the outer shock
(Fig.\,\ref{fig:2D_cont}d).  Like the outer shock, the inner shock
eventually becomes prolate, too.
\begin{figure}[h]
\epsfxsize=8.8cm \epsfclipon \epsffile{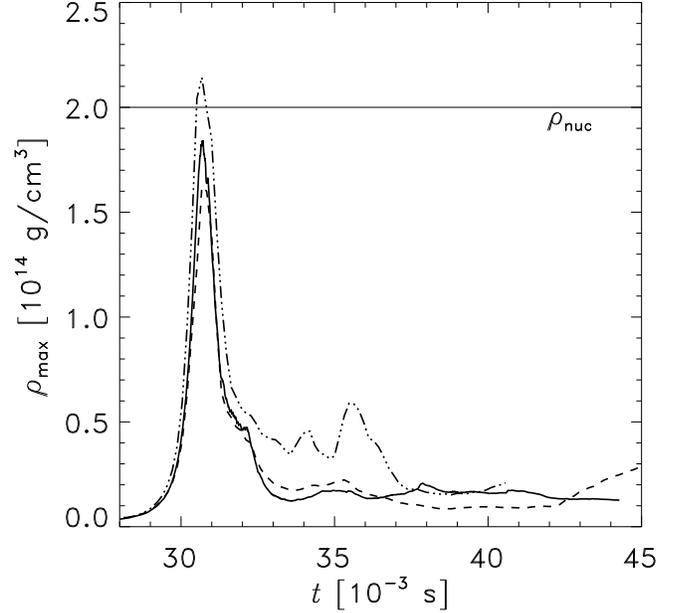}
\caption[]{Maximum density on the grid as a function of time for the
three-dimensional models MD1 (dashed) and MD2 (dashed-dotted).  For
comparison we also show the maximum density of the two-dimensional model
A4B5G5 (solid).  The solid horizontal line marks the assumed nuclear
matter density $\varrho_{\rm nuc} = 2\,10^{14}\gcc$.}
\label{fig:3Drh}
\end{figure}

The overall dynamics of the core can be conveniently analyzed in terms
of the radial extent of different ``mass shells'' $M(R)$ defined by

\beq
\label{m_r}
M(R) := 2\pi\int_0^R\int_0^\pi
       \langle \varrho(r,\theta) \rangle_\phi\,r^2\sin\theta
       \,{\rm d}\theta{\rm d} r  \,,
\eeq

where $\langle \rangle_\phi$ denotes the azimuthal mean of the density
to be calculated in the 3D simulations.  $M(R)$ is the mass contained in
a sphere of radius $R$.  Note that in more than one spatial dimension
these mass shells are not identical with the well known Lagrangian mass
shells.

Inverting $M(R)$ one obtains the radius $R_M$ of the mass shell.  This
quantity is plotted as a function of time for different masses $M$ in
Fig.\,\ref{fig:m_r}.  One recognizes that maximum compression is
reached in model A4B5G5 at $t=30.67$\,ms, which is followed by an
expansion lasting for about 2\,ms.  Without any further significant
oscillations the inner core evolves towards its new equilibrium: The
radii of mass shells with $M\la 0.7\ms$ are nearly constant with time
for $t\ga 33$\,ms (Fig.\,\ref{fig:m_r}).  The same behavior can be
deduced from the evolution of the rotation parameter ($\beta\propto
R^{-1}$ for a homogeneous sphere with radius $R$) and of the maximum
density on the grid (Figs.\,\ref{fig:3Dbeta} and \ref{fig:3Drh}).
Figure\,\ref{fig:m_r} also shows that the central core $M(R) < 0.7\ms$
oscillates with a single dominant volume mode, which is in agreement
with the results obtained by M\"onch\-meyer \etal (1991).  They found
single volume modes in cores with coherent motion in equatorial and
polar directions.  This certainly occurs in model A4B5G5, since matter
in the torus which contains most of the mass shows almost no motion in
polar directions at all.

\begin{figure*}[t]
 \begin{tabular}{cc}
\put(0.9,0.3){{\LARGE\bf a}}
\epsfxsize=8.8cm \epsfclipon \epsffile{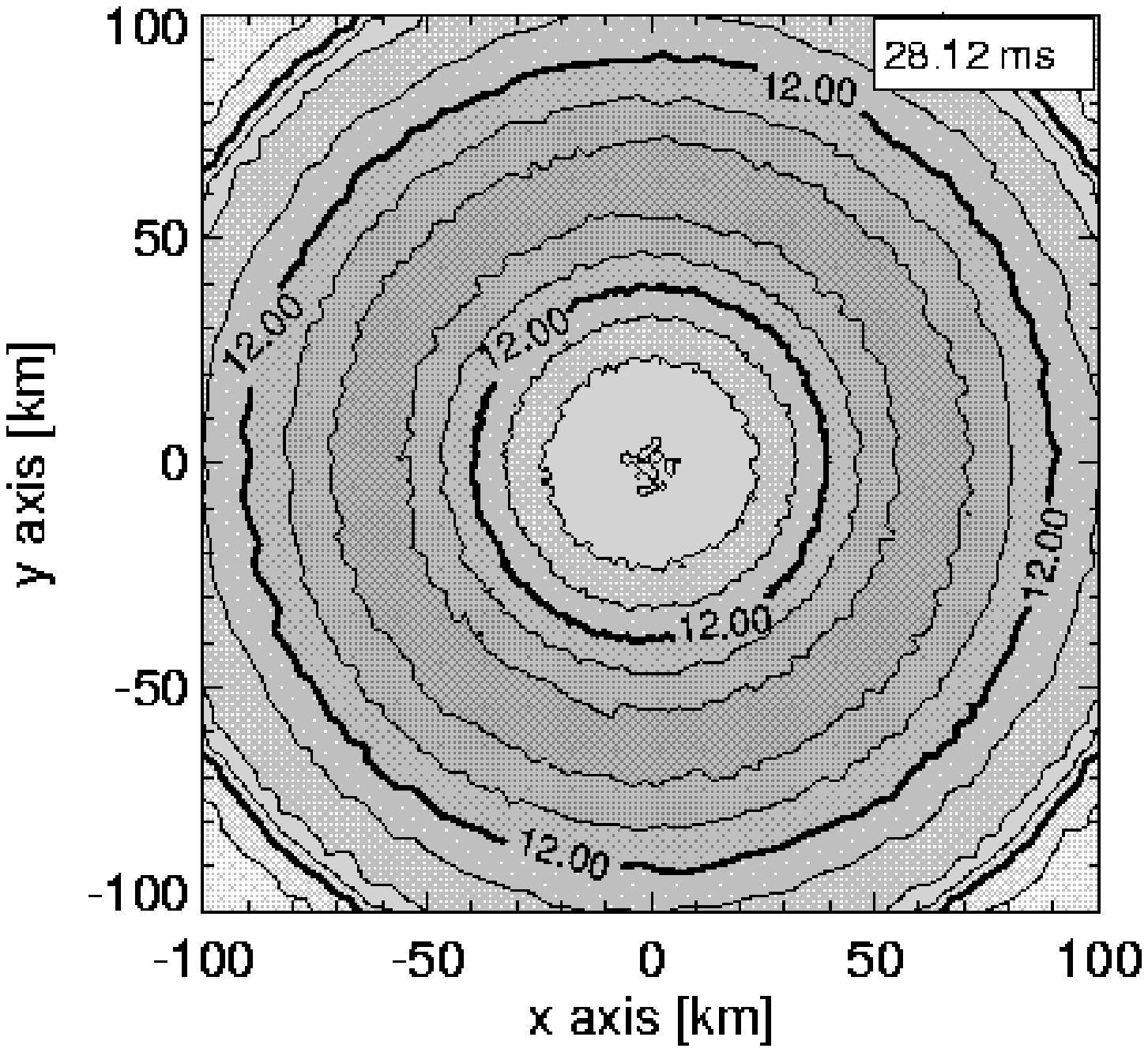} &
\put(0.9,0.3){{\LARGE\bf b}}
\epsfxsize=8.8cm \epsfclipon \epsffile{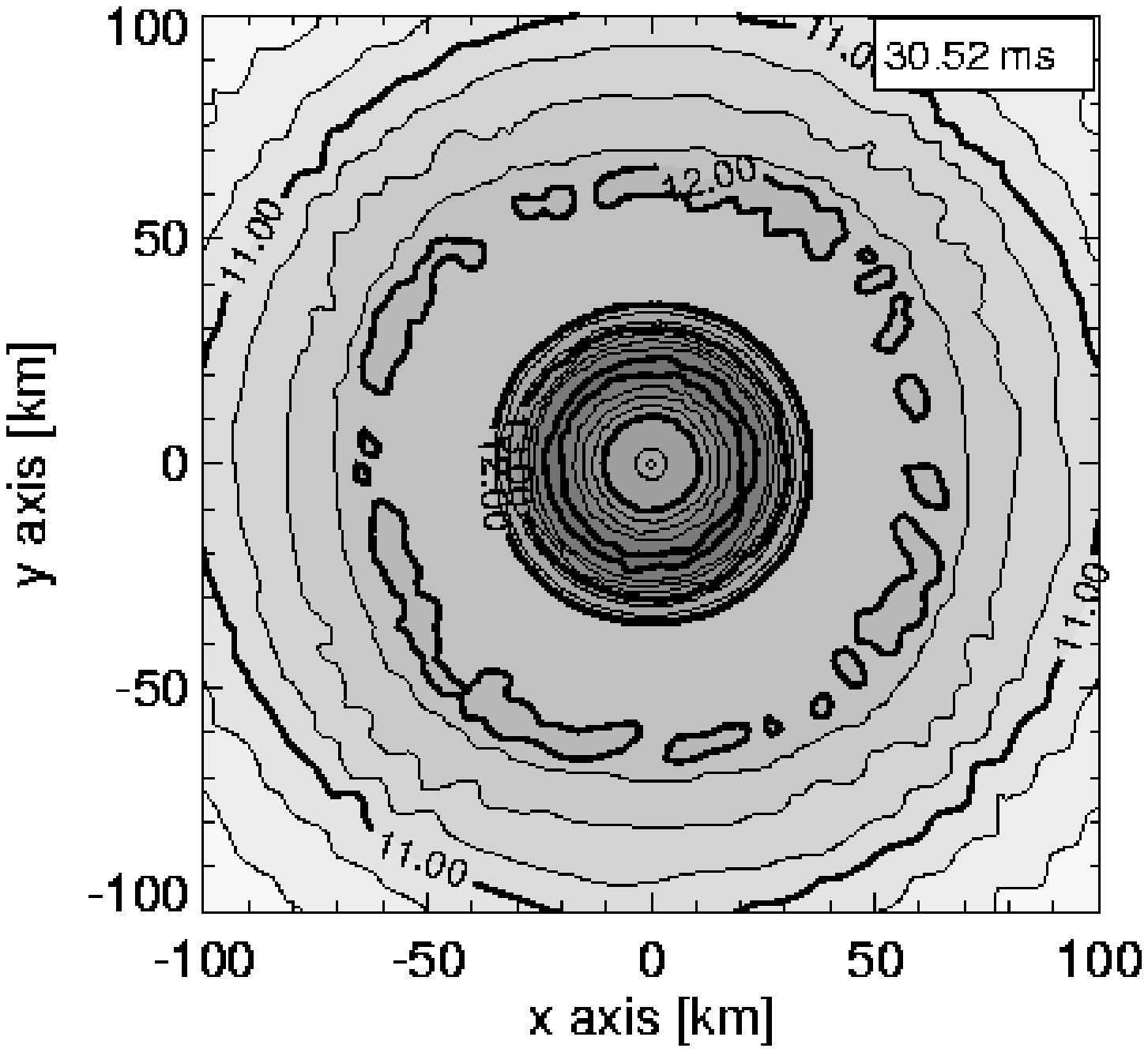} \\[-2ex]
\put(0.9,0.3){{\LARGE\bf c}}
\epsfxsize=8.8cm \epsfclipon \epsffile{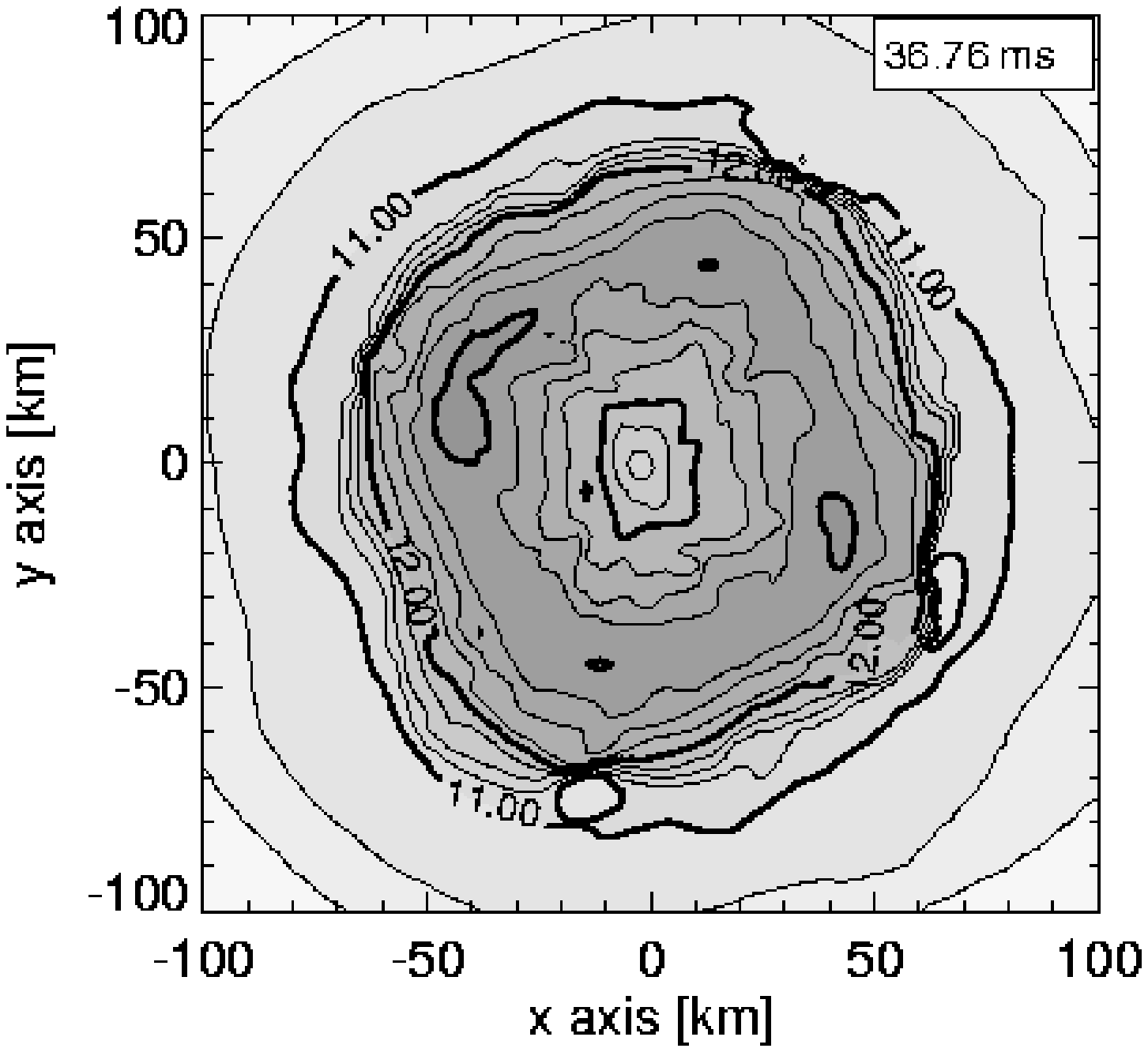} &
\put(0.9,0.3){{\LARGE\bf d}}
\epsfxsize=8.8cm \epsfclipon \epsffile{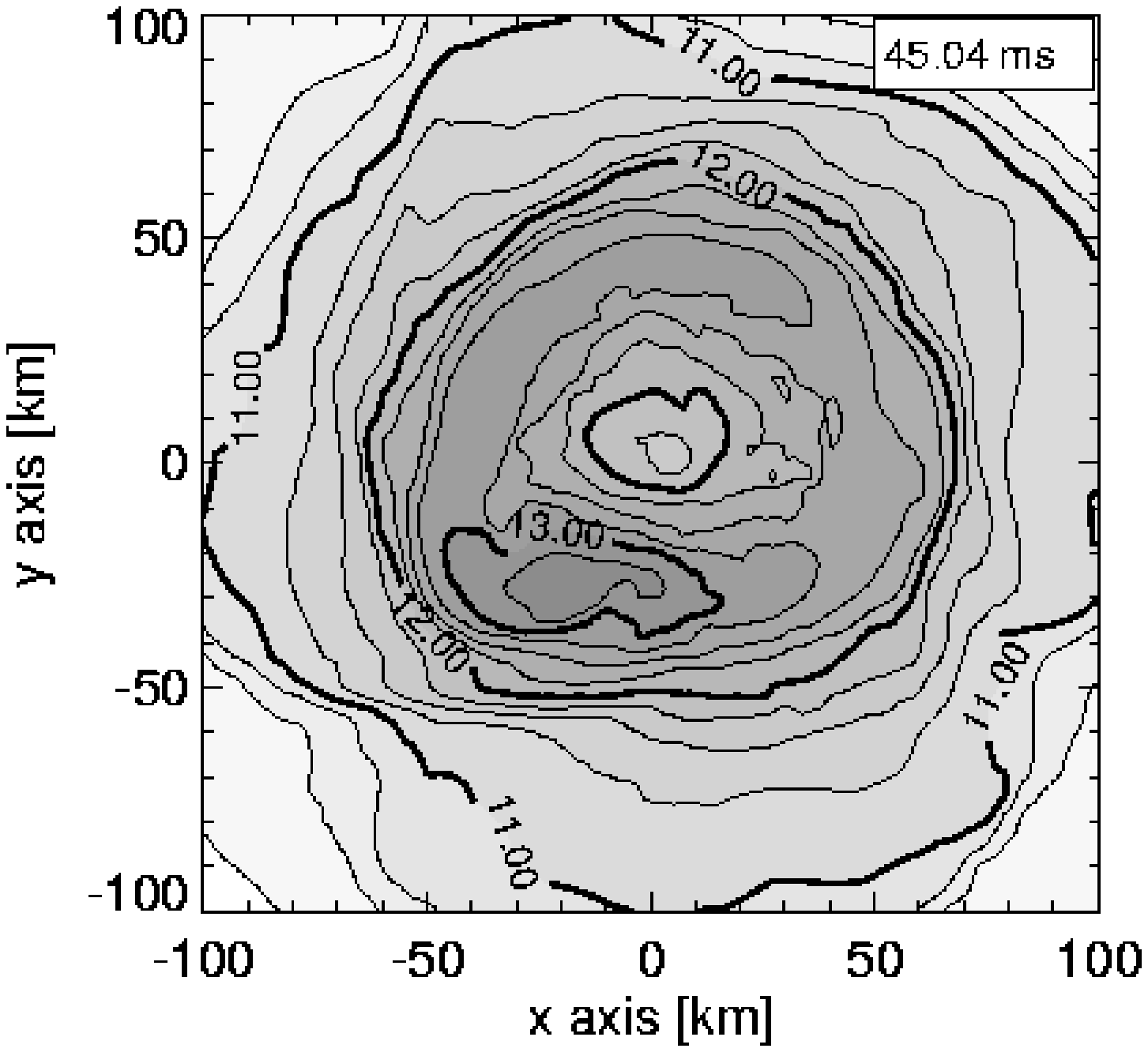} \\[-2ex]
 \end{tabular}
\caption[]{Snapshots of the density distribution (in units of
[$\gcc$]) in the equatorial plane of model MD1. The contours are
logarithmically spaced with intervals of 0.25 dex, they are shaded
with darker grey values for higher density regions and labeled with
their respective values. The time of the snapshot is given in the
upper right corner of each panel.}
\label{fig:3Dcont1}
\end{figure*}

Matter in the torus (about $0.7\ms$) stays in sonic contact, \ie it
contracts coherently.  It also remains unshocked well beyond core
bounce.  Hence, the torus resembles the qualitative features of a
homologously contracting inner core that has been found analytically for
the spherical collapse of polytropic cores (Goldreich \& Weber 1980,
Yahil 1983) and numerically in slowly rotating cores with ellipsoidal
density stratification (Finn \& Evans 1990; M\"onch\-meyer \etal
1991).

The main features of the quadrupole amplitude $A^{\rm E2}_{20}$
calculated for model A4B5G5 are:  A slow monotonic increase with time
for the first 25\,ms is followed by a pronounced negative spike at the
time of bounce with no subsequent short-period oscillations (see
Fig.\,\ref{fig:3Dh}).  We therefore consider the gravitational waveform
to be closest resembled by signal type\,II, which is characterized by
prominent spikes arising at the bounce(s) due to single dominant volume
mode(s) (M\"onch\-meyer \etal 1991). 
 
\noindent 
The non-vanishing polarization, calculated for a source at 10\,Mpc and
for an observer located in the equatorial plane of the core, has a
peak value of $h_+ = -3.5\,10^{-23}$.
\begin{sloppy}
The energy of the quadrupole radiation is $E_{\rm GW}= 7.8\,10^{-8}
\ms\,c^2$.  Most of the spectral power is radiated at a frequency of
$\nu \approx 200$\,Hz.  This is the largest signal strength obtained
for any axisymmetric collapse models in the set of Zwerger \& M\"uller
(1997).  However, the amplitude is still by at least one order of
magnitude too small to be detectable even with the ``advanced'' LIGO
interferometer (see Abramovici \etal 1992 for the sensitivity limits),
if the source is located in the Virgo cluster.
\end{sloppy}

\subsection{Evolution of non-axisymmetric models}

\begin{figure*}[p]
 \begin{tabular}{cc}
\put(0.9,0.3){{\LARGE\bf a}}
 \epsfxsize=8.8cm \epsfclipon \epsffile{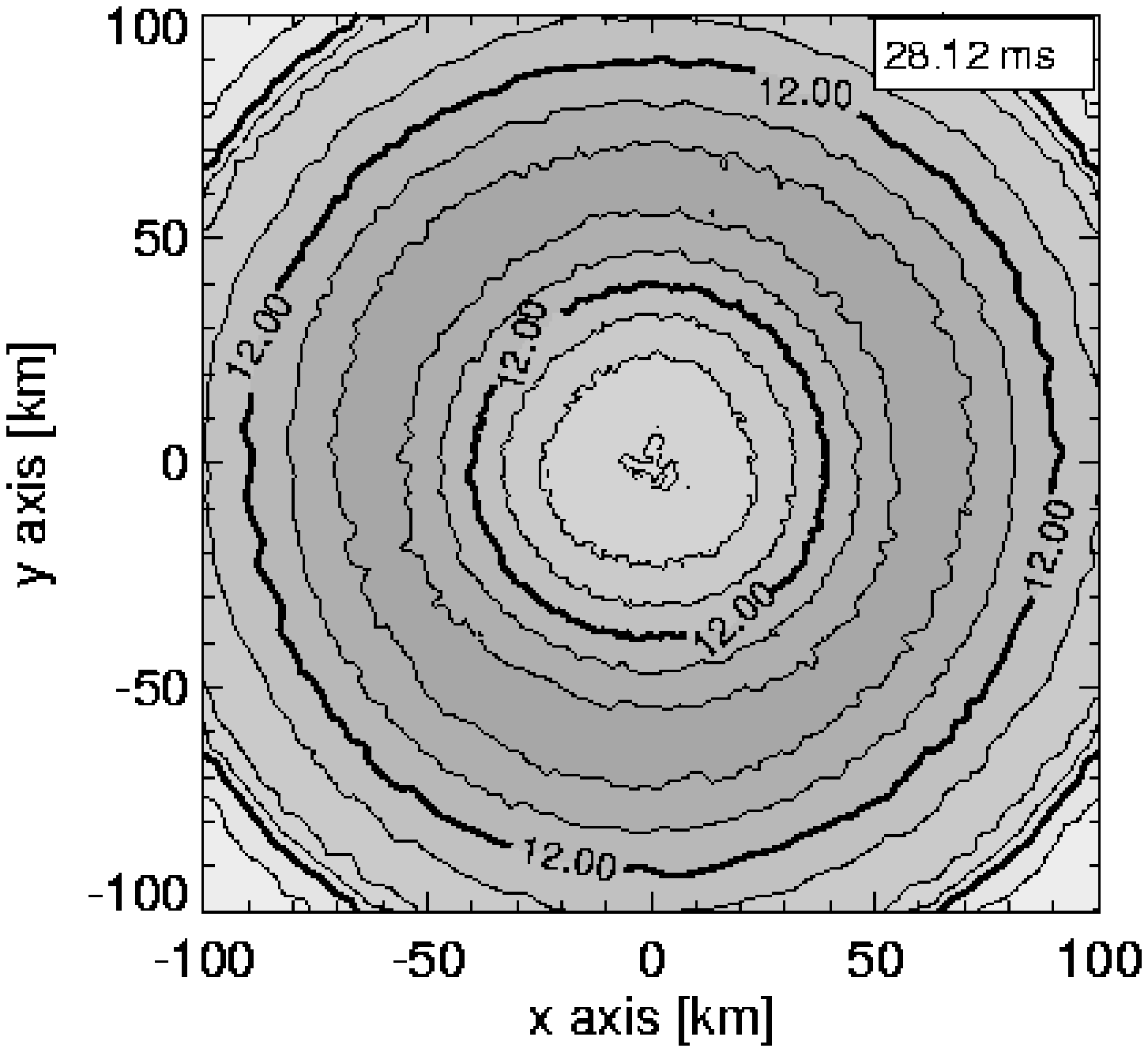} &
\put(0.9,0.3){{\LARGE\bf b}}
 \epsfxsize=8.8cm \epsfclipon \epsffile{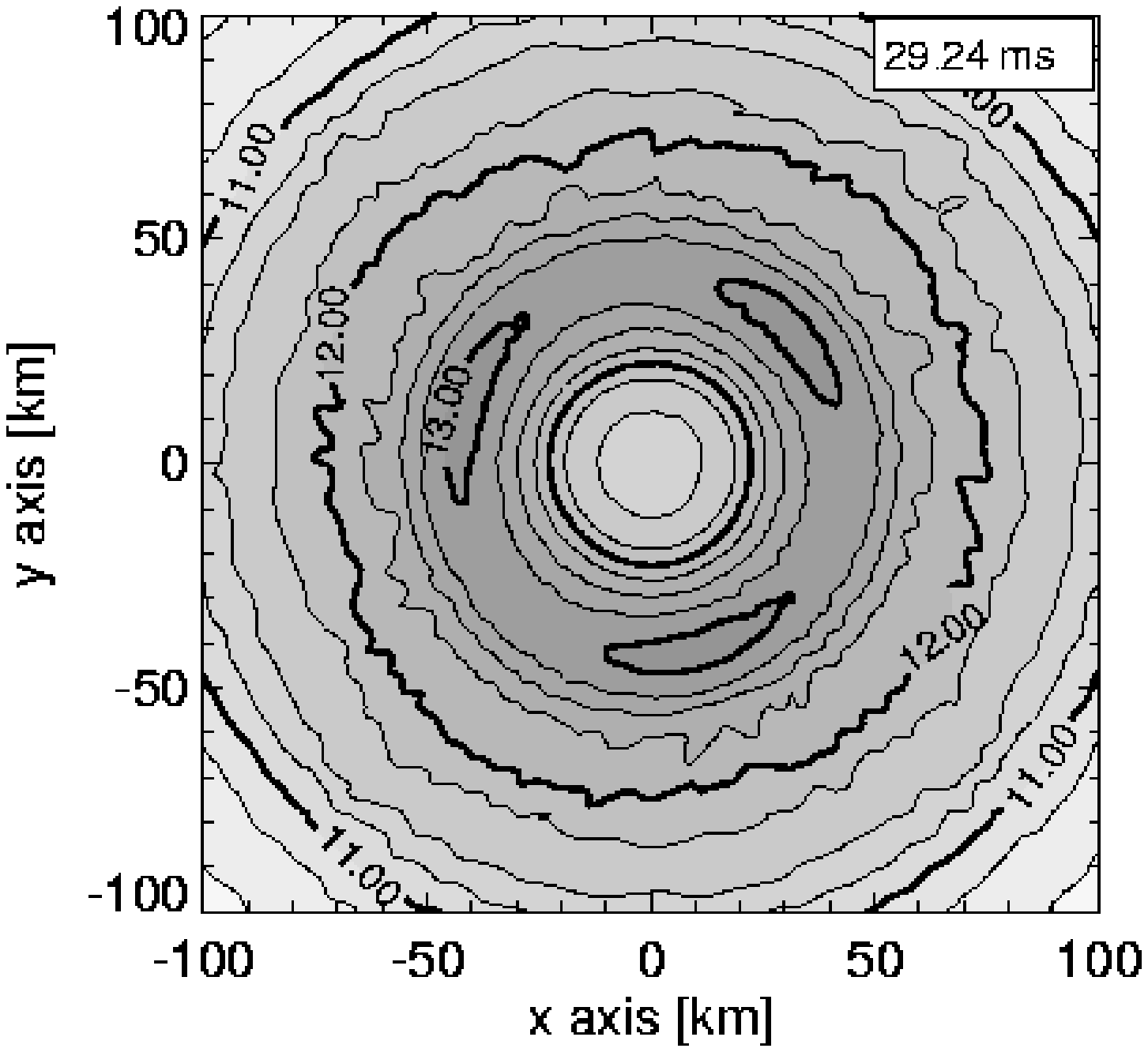} \\[-2ex]
\put(0.9,0.3){{\LARGE\bf c}}
 \epsfxsize=8.8cm \epsfclipon \epsffile{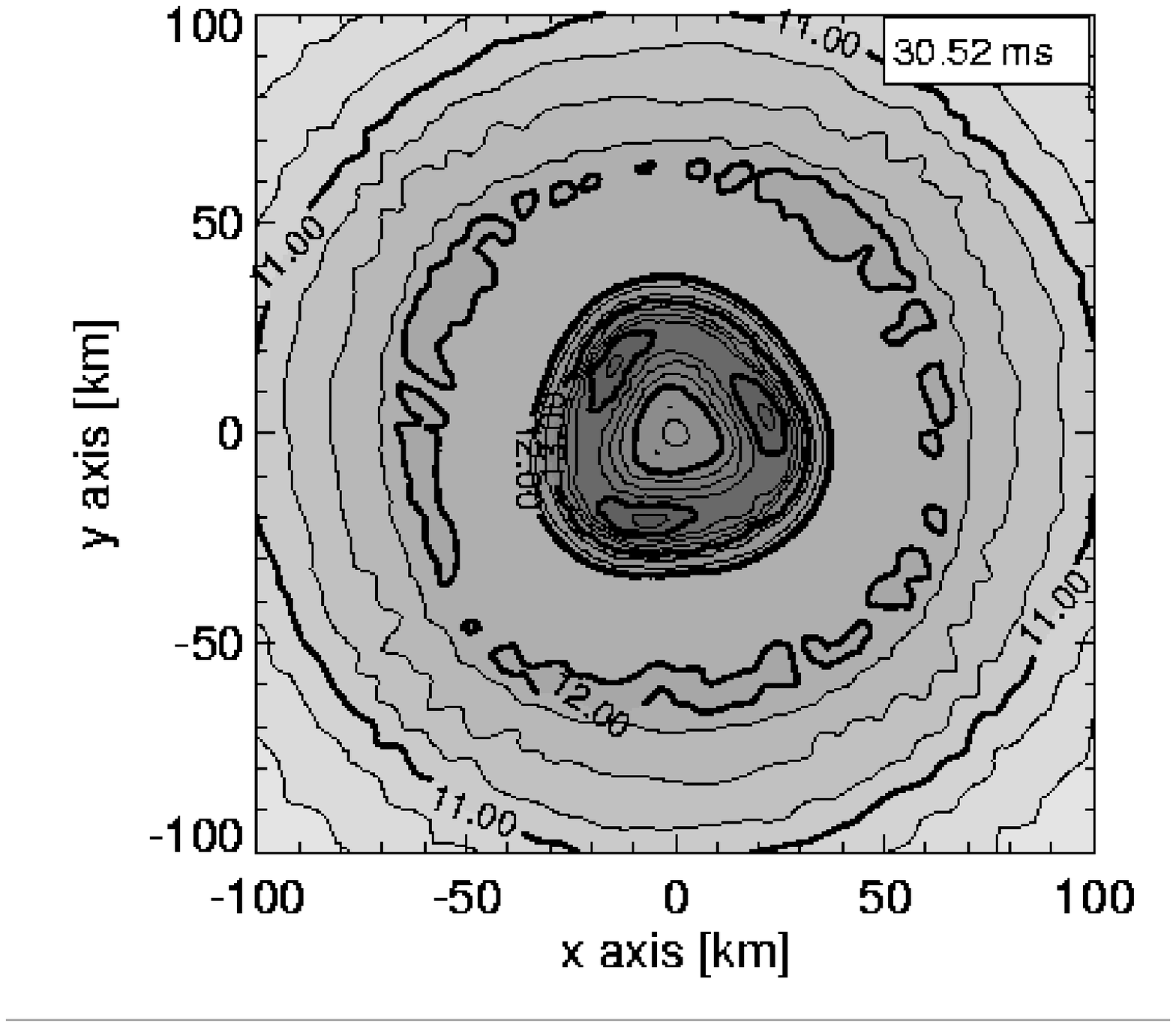} &
\put(0.9,0.3){{\LARGE\bf d}}
 \epsfxsize=8.8cm \epsfclipon \epsffile{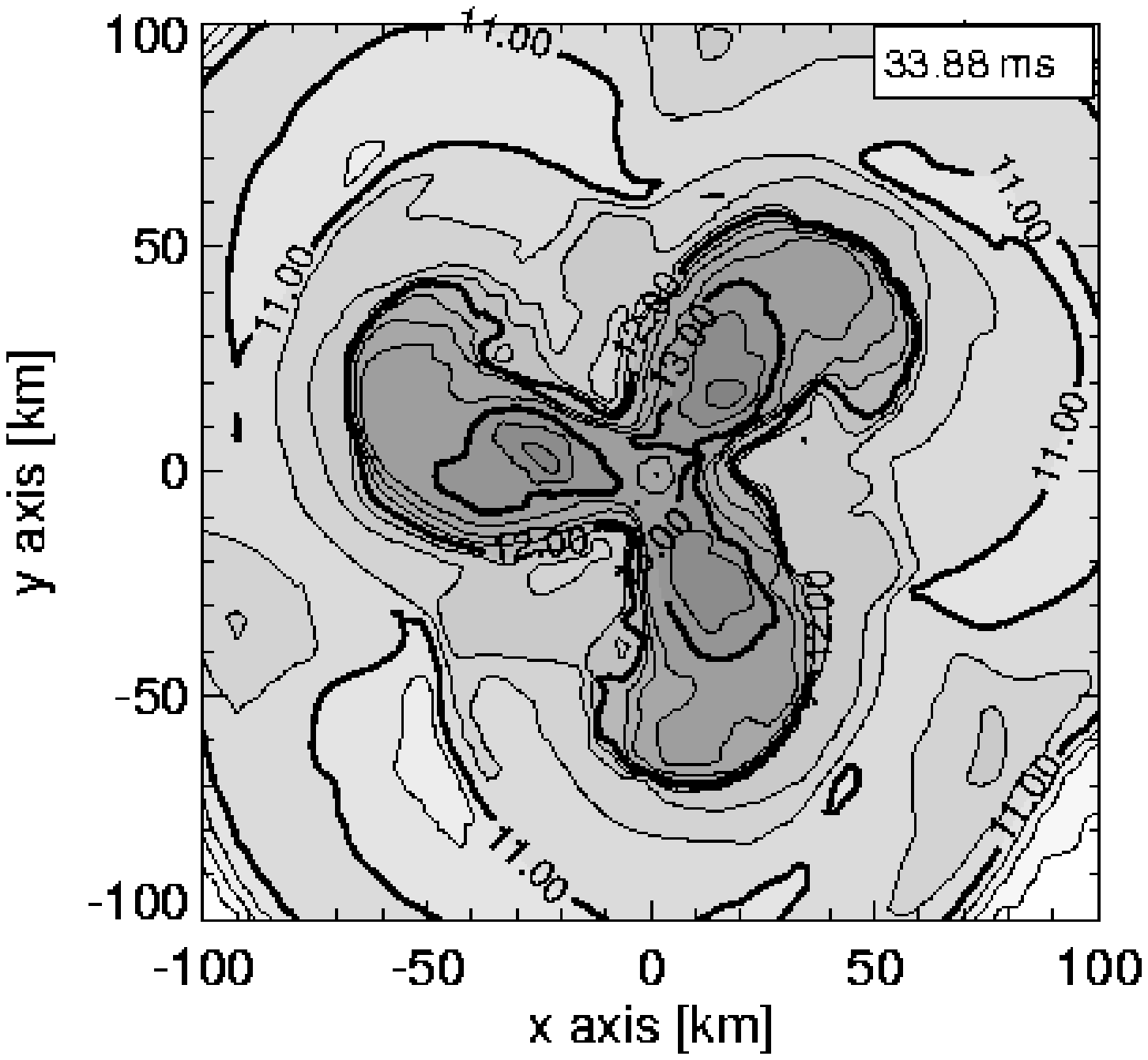} \\[-2ex]
\put(0.9,0.3){{\LARGE\bf e}}
 \epsfxsize=8.8cm \epsfclipon \epsffile{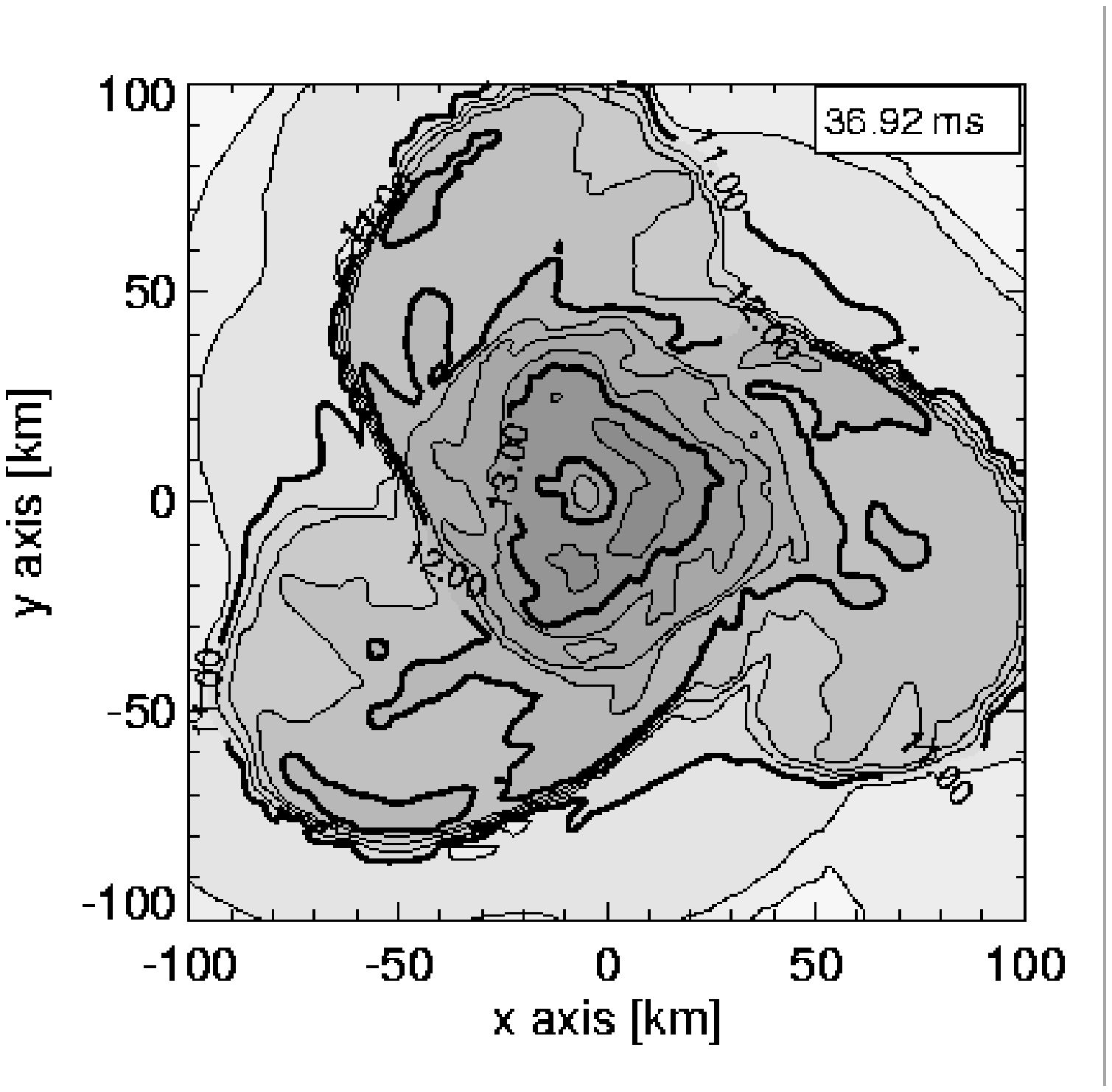} &
\put(0.9,0.3){{\LARGE\bf f}}
 \epsfxsize=8.8cm \epsfclipon \epsffile{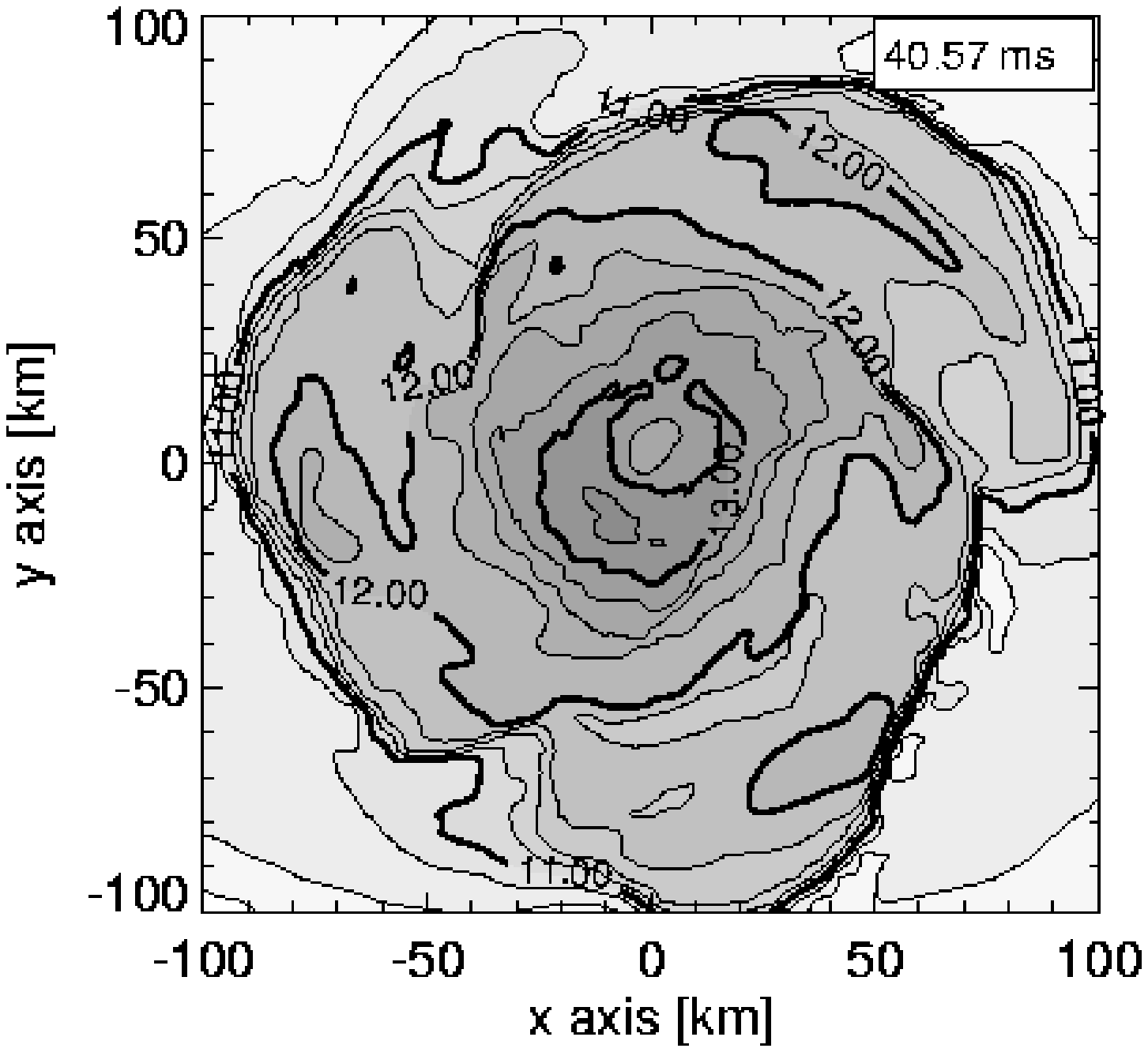} \\[-2ex]
 \end{tabular}
\caption[]{Snapshots of the density distribution (in units of
[$\gcc$]) in the equatorial plane of model MD2. The contours are
logarithmically spaced with intervals of 0.25 dex, they are shaded
with darker grey values for higher density regions and labeled with
their respective values. The time of each snapshot is given in the
upper right corner of each panel.}
\label{fig:3Dcont2}
\end{figure*}

The initial non-axisymmetric perturbations were introduced at
$t_0=28.13$\,ms (Figs.\,\ref{fig:3Dcont1}a and \ref{fig:3Dcont2}a).
This is about 2.5\,ms before bounce (when the density reaches a
maximum inside the torus), and about 2\,ms before the core's rotation
rates $\beta \ga 0.3$, \ie when it should become dynamically unstable.
The additional perturbation of 5\% amplitude and $m=3$ azimuthal
dependence imposed on model MD2 can hardly be distinguished from the
10\% random noise perturbations of both models.  The inner
torus has contracted from an initial radius of 200\,km to one of
60\,km at $t=t_0$. Note that we refer to its density maximum, when we
give its radial position.

The subsequent rapid contraction of the rotating core is reflected by
a steep rise of the maximum density, which peaks at $t=30.68$\,ms in
both 3D models and in the axisymmetric one (Fig.\,\ref{fig:3Drh}).  At
approximately the same time the contraction (Fig.\,\ref{fig:m_r}), the
rotation parameter $\beta$ (Fig.\,\ref{fig:3Dbeta}), the
dissipation-rate of kinetic-infall energy into thermal energy and the
gravitational potential energy reach their peak values.  Thus, we
consider $t=30.68$\,ms to be the time of bounce, although the {\em
central} density does not reach its first peak then, which is usually
considered as the bounce criterion. However, because of the torus-like
density stratification of the models, the central density has less
meaning. During the simulations it remained always much lower than the
maximum density inside the torus (by 1 -- 3 orders of magnitude).  One
also has to be careful, in particular in 3D simulations, when deriving
implications from ``local'' quantities. For example, one might be
tempted to conclude from the fact that the maximum density exceeds
nuclear matter density in model MD2, \ie $\varrho_{\rm
max}>2\,10^{14}\gcc$ (Fig.\,\ref{fig:3Drh}), that this model contrary
to model MD1 suffered a bounce due to the stiffening of the equation
of state.  However, nuclear matter density is exceeded only in a few
zones inside the torus (see Fig.\,\ref{fig:3Dcont2}c), while the
corresponding azimuthal average, which is determining the overall
dynamics, is practically identical to that in model MD1.  Therefore,
we always compare the value of the maximum density with the
corresponding azimuthal mean, before drawing any conclusions.

Up to $t\approx 33$\,ms, \ie up to about 2\,ms after bounce, the
non-axisymmetric initial perturbations have not grown far enough to
produce any significant changes in the overall dynamics of the
collapsing core. This can be seen by comparing the time evolution of
the radii of mass shells defined by Eq.\,(\ref{m_r}) (see
Fig.\,\ref{fig:m_r}).  Also locally, non-axisym\-metries are still small
perturbations on a rapidly changing axisymmetric background. The
density contrast in the torus is $\delta \varrho\la 0.15$ for model MD2
at $t\le 33$\,ms (Figs.\,\ref{fig:3Dcont2}a--c).  Hence, the
distribution of hydrodynamic variables in planes perpendicular to the
rotation axis is still given by
Fig.\,\ref{fig:2D_cont}.

\begin{figure}[!t]
\begin{tabular}{cc}
\epsfxsize=4.2cm \epsfclipon \epsffile{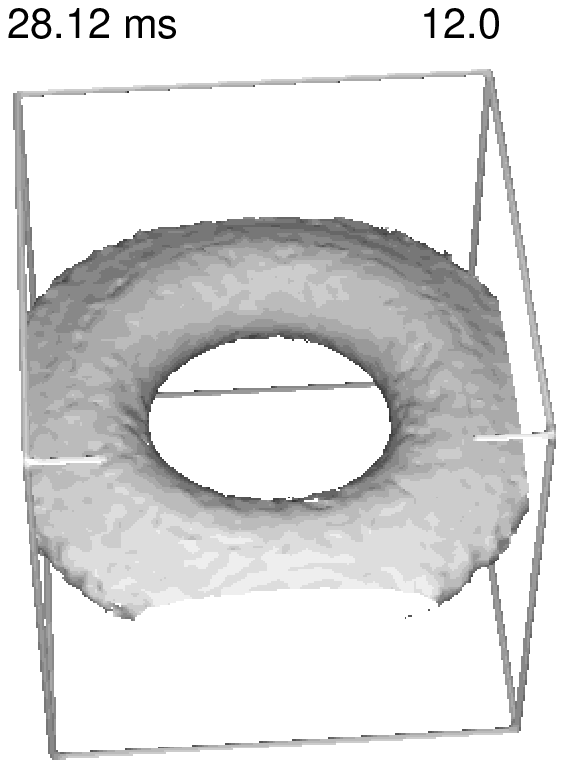} &
\epsfxsize=4.2cm \epsfclipon \epsffile{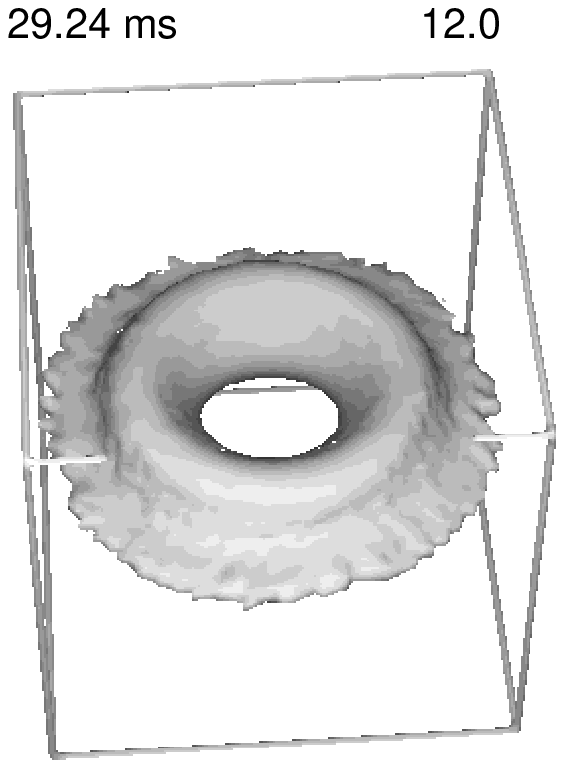} \\ [-2ex]
\epsfxsize=4.2cm \epsfclipon \epsffile{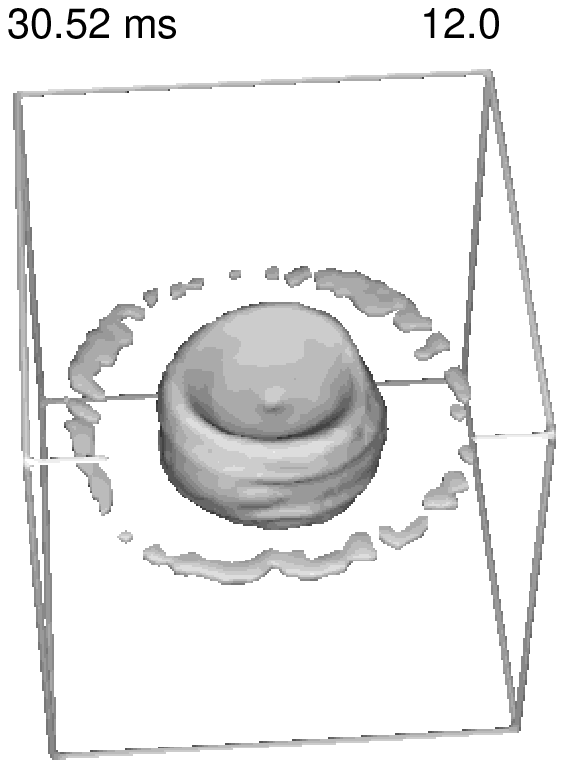} &
\epsfxsize=4.2cm \epsfclipon \epsffile{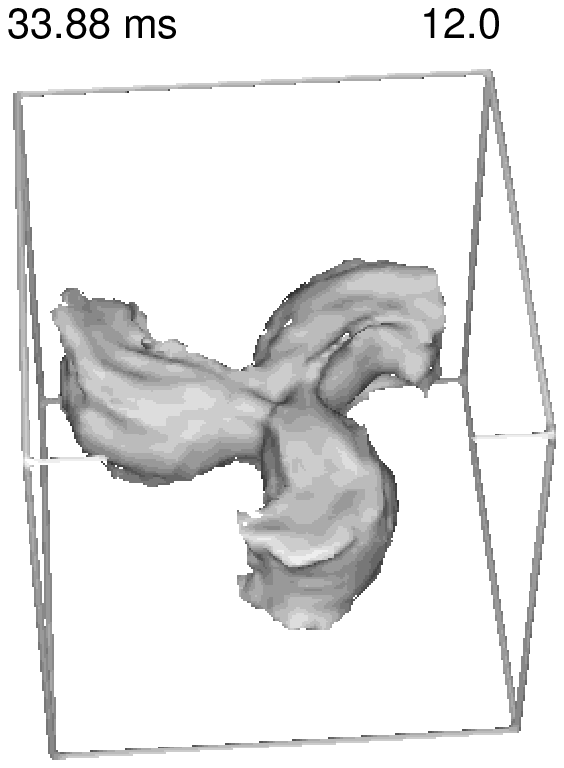} \\ [-2ex]
\epsfxsize=4.2cm \epsfclipon \epsffile{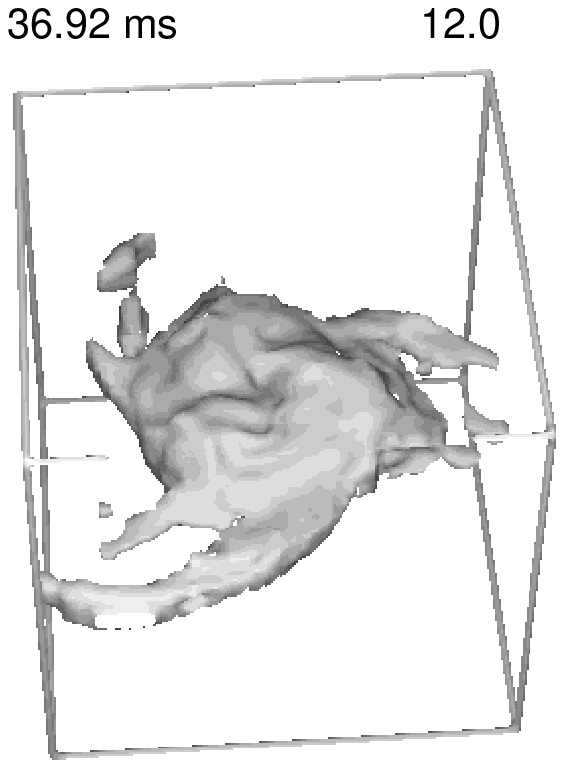} &
\epsfxsize=4.2cm \epsfclipon \epsffile{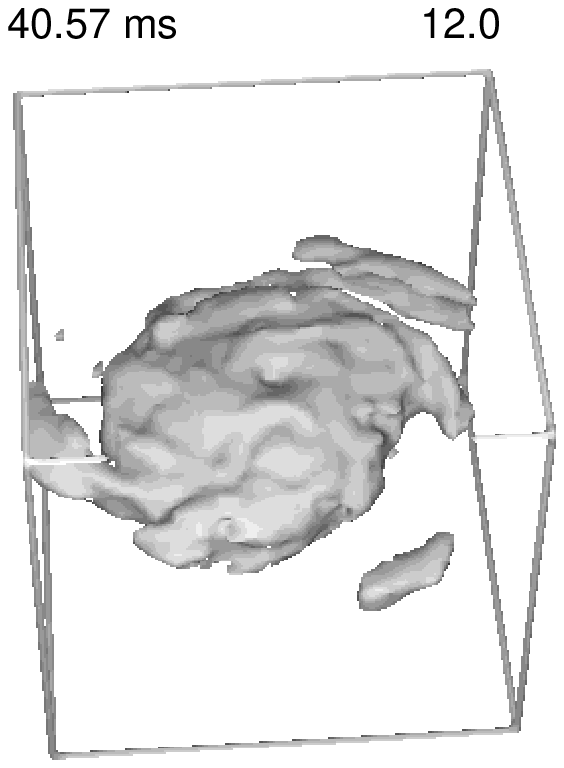} \\ [-2ex]
\end{tabular}
\caption{Surfaces of constant density $\varrho =10^{12}\gcc$ for model
MD2. The edges of the cubic box have a length of 160\,km. The
snapshots are taken at the same times as those in
Fig.\,\protect\ref{fig:3Dcont2}.}
\label{fig:LA38V}
\end{figure}

The subsequent evolution differs from that of the axisymmetric model
and is also different between the two 3D models.

{\em Model MD1:} Due to the Cartesian geometry of the computational
grid, the toroidal mode with $m=4$ sticks out of the noise first. The
inner torus assumes a transient square-like structure, which exists
for $\approx 7$\,ms. This does not influence the overall dynamics of
the model significantly compared to the axisymmetric one, although we
find a somewhat less compact density stratification when comparing the
radii of the innermost mass shells with those of model A4B5G5 for $t
\ga 33$\,ms (Fig.\,\ref{fig:m_r}).  When we stopped the simulation at
$t=45.04$\,ms, the final configuration shows a prominent off-center
density maximum in the equatorial plane rotating with a period of
$T\approx 5$\,ms at a distance of $\varpi \approx 30$\,km.

{\em Model MD2}: The development of the instability is \linebreak
shown in Figs.\,\ref{fig:3Dcont1} and \ref{fig:LA38V}.  By the time of
bounce three distinct density maxima are visible the density contrast
being $\delta \varrho \la 0.15$ inside the torus. These density maxima
eventually grow into three distinct clumps (Fig.\,\ref{fig:3Dcont2}c).
The further evolution is characterized by intensive hydrodynamic
activity produced by the twirling-stick action of the three clumps.
Most notably are trailing ``spiral arms'' causing mass and angular
momentum to be transported away from the non-axisymmetric, high
density regions (Figs.\,\linebreak \ref{fig:3Dcont2}d, e). This is
mirrored in the evolution of the innermost ($M \le
0.7\ms$) mass shells, whose radii decrease during the time when the
spiral arms are present. The largest deviations from the monotonic
expansion of the mass shells of model MD1 occur between $t \approx
33$\,ms and $t \approx 38$\,ms, when the spiral arms are most
prominent in model MD2.  Eventually, the three spiral arms merge
(Fig.\,\ref{fig:LA38V} second last
snapshot and Fig.\,\ref{fig:3Dcont2}e) and form a bar-like object inside the $M=0.7\ms$ mass shell
(Fig.\,\ref{fig:LA38V} last snapshot and Fig.\,\ref{fig:3Dcont2}f).

\subsection{Growth region of the instability}

\begin{figure}[!t]
\epsfxsize=8.8 cm \epsfclipon \epsffile{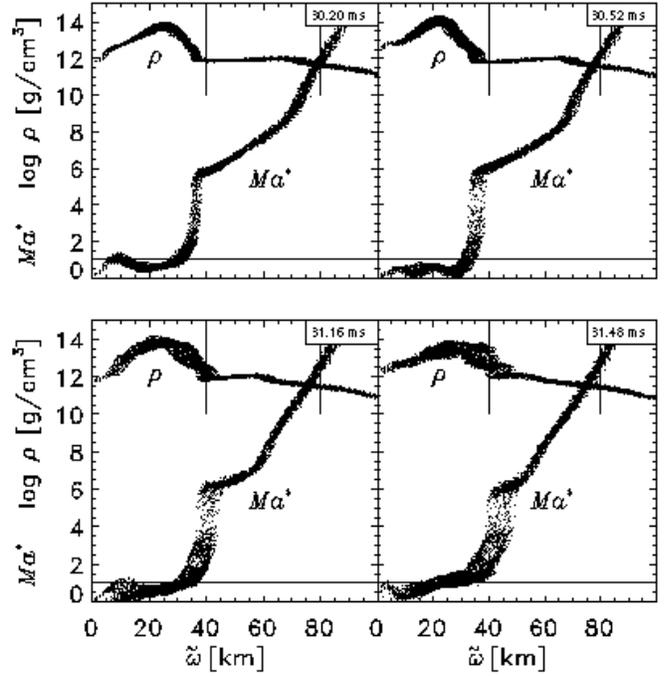}
\caption[]{Scatter plots of the density (upper ``curves'', logarithmic
scale, in units of $[\gcc]$) and the Mach number $\Ma$ (lower
``curves'', calculated in a coordinate system rigidly rotating with
the torus) in the equatorial plane $z\teq 0$ versus distance
$\varpi:=\sqrt{x^2+y^2}$ from the rotation axis at different instants
in time (see upper right corner of each panel). The bold horizontal
line separates supersonic (upper) from subsonic (lower) flow regions.
The thin vertical lines indicate the boundaries $\xi_{\rm bound}$ of
the nested grids (see text).}
\label{fig:mach}
\end{figure}

For both 3D models we find that although perturbations have been
imposed on the whole computational grid with the same relative
amplitude initially, they grow only in the immediate vicinity of the
inner torus, \ie inside a cube with an edge length of $\approx
100$\,km (Figs.\,\ref{fig:LA38V} and \ref{fig:3Dcont2}).  This can be
seen more clearly in Fig.\,\ref{fig:mach}, where we have plotted the
density (upper ``curves'') of each grid cell in the equatorial plane
versus its (cylindrical) coordinate distance $\varpi:=\sqrt{x^2+y^2}$
from the origin. One notices considerable deviations from axisymmetry
(which is mirrored in Fig.\,\ref{fig:mach} by a large spread in
density at a given distance $\varpi$) only for $\varpi\la
40\,\rm{km}$.  That this is not an artefact due to the usage of a
multiple-nested refined grid can be seen as follows: The cube-shaped
boundaries of the single grids are located at a {\em Cartesian}
coordinate distance $\xi_{\rm bound}\in\{40\,\rm{km}, 80\,\rm{km},
160\,\rm{km},\dots\}$ from the origin ($\xi\in\{x,y,z\}$).  Depending
on the azimuthal angle $\phi$ this implies for each grid a doubling of
the (linear) zone size at {\em cylindrical} coordinate distances
$\varpi_{\rm bound}\in[\xi_{\rm bound},(1+\sqrt{2})\cdot \xi_{\rm
bound}]$.  If the growth of the perturbations was damped notably by
the coarser numerical resolution at larger radii one would expect the
spread in density as function of $\varpi$ to change significantly at
the limiting values of $\varpi_{\rm bound}$, which is obviously not
the case in Fig.\,\ref{fig:mach}.

In contrast, we observe a striking correlation between the Mach-number
$\Ma$ calculated in a rotating coordinate system whose angular
velocity $\Omega \approx 1800 {\rm s}^{-1}$ is that of the torus at $t=30$\,ms (when $\beta = \beta_{\rm
dyn}$) and the
domain where the density varies considerably with azimuthal angle:
Non-axisymmetric perturbations grow only in regions where
the flow is subsonic, \ie where $\Ma < 1$ (Fig.\,\ref{fig:mach}).

Our qualitative interpretation of this result is as follows:
Non-axisymmetric instabilities have been shown to occur in MacLaurin
spheroids as well as in a large variety of (differentially) rotating
compressible fluid bodies {\em in equilibrium}, if the rate of
rotation is sufficiently large.  However, being equilibrium
configurations, the instabilities in these models grow in the absence
of any radial motion. When viewed from a suitably chosen rotating
coordinate system sonic contact along the azimuthal coordinate is
established in these models and allows for a coherent growth of the
instability.  According to our results, the latter condition seems to
be {\em sufficient} for the growth of non-axisymmetric instabilities
also in collapsing rotators.  If on the other hand sonic contact is
not given along the azimuthal coordinate due to large radial
velocities --- like in the outer core of our models --- it is
difficult to imagine how global tri-axial deformations can develop
coherently.  Thus, we suppose that sonic communication in azimuthal
direction is also a {\em necessary} condition for the growth of
non-axisymmetric instabilities.

One might argue that these results depend on our particular choice of
the rotating coordinate system. However, establishing co-rotation at
any distance $\varpi$ (\ie chosing a coordinate system rotating with
an angular velocity equal to that of the fluid at that distance)
outside the domain of the inner torus does not allow to ``transform
away'' supersonic velocities in the collapsing (outer) core, since the
radial component dominates the angular component of the velocity field
by a large amount.

\subsection{Gravitational wave signal}

\begin{figure}[!t]
\epsfxsize=8.8cm \epsfclipon \epsffile{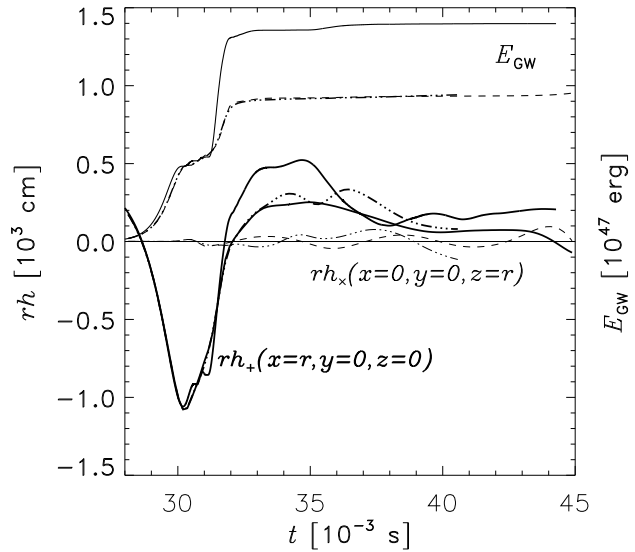}
\caption[]{Gravitational wave amplitudes $rh_+(x\teq r,y\teq 0,z\teq
0)$, (thick lines), $rh_{\times}(x\teq 0,y\teq 0,z\teq r)$, (thin
lines) and radiated energy $E_{\rm GW}$ as functions of time for the
three-dimensional models MD1 (dashed) and MD2 (dashed-dotted) in
comparison with the axisymmetric calculation (solid). The peak values
$|rh| \simeq 1000$\,cm correspond to dimensionless amplitudes $|h|
\approx 3\,10^{-23}$ for a source at $r=10$\,Mpc.}
\label{fig:3Dh}
\end{figure}

Although models MD1 and MD2 show prominent deviations from
axisymmetry, the gravitational wave signal is changed only marginally
compared with the axisymmetric calculation. In particular, the maximum
wave amplitudes are equal for the 3D and the 2D simulations within an
accuracy of 2\%.  The waveforms and the energy radiated in form of
gravitational waves are displayed in Fig.\,\ref{fig:3Dh}. The peak
amplitude of $h_+$ is reached at about the time of bounce in both the
2D and the 3D models. Up to this point in the evolution the waveforms
are identical, too. This is not surprising, since the initial
perturbations of models MD1 and MD2 have not grown significantly until
the time of bounce, which occurs only 0.5\,ms after $\beta_{\rm dyn}$
is reached. As discussed above, the density contrast is $\delta
\varrho\la 0.15$ inside the inner torus before bounce.  Accordingly,
$h_\times$, which is of genuine non-axisymmetric origin nearly
vanishes until bounce (Fig.\,\ref{fig:3Dh}).  

The maximum amplitudes of $h_\times$ are reached during the further
evolution, when the spiral arms merge to form the final bar (model
MD2) or when the transient $m=4$ structure evolves to some
``bar-like'' configuration (model MD1). These maxima are only of the
order of 10\% of the maximum value of $|h_+|$.  The amplitudes of the
cross- and plus-polarizations finally become comparable, because the
inner core (torus) approaches its new rotational equilibrium and thus
the time derivatives of the quadrupole moments due to radial motion
become steadily smaller. The periodicity seen in $h_\times$ is due to
a nearly solid-body type rotation of the high density regions of the
core.

Comparing the 3D results with the 2D signal, one notices that the 3D
waveforms do not reproduce the small additional local minimum in $h_+$
visible at $t \approx 31$\,ms for the axisymmetric model. Also the
(absolute values of the) time derivatives of $h_+$ are smaller for the
following 2\,ms. This leads to a smaller value of the amount of energy
radiated in form of gravitational waves (Eq.\,\ref{E_GW}; for model
MD1 $h_+$ is roughly proportional to $\ddot I {\bari}_{xx}$ and $\ddot
I {\bari}_{yy}$ during this time interval because of the near
axisymmetry of the core at this time).  We have analyzed carefully the
unexpected fact that compared to the axisymmetric models we obtain
only 65\% of the gravitational wave energy for the two models MD1 and
MD2, which during this epoch are approximately, but not exactly
axisymmetric.  Several tests and comparisons (\eg with poorer resolved
2D calculations) have shown that this is {\em not} a possibly
unnoticed 3D effect, but due to a somewhat lower ``angular''
resolution of the 3D simulation compared with the best resolved 2D run
(see Rampp (1997) for details). The lower resolution gives rise to a
larger violation of total energy conservation, most of which occurs
during bounce.  Since all dynamical quantities of model MD1 -- local
as well as global ones -- agree with the 2D results within an accuracy
of a few percent, we consider the observationally relevant {\em
waveforms} calculated in the 3D simulations to be reliable within the
underlying approximations (cf.~sect.~2).  We also point out the fact
that the squared time derivatives of the moments $\ddot I {\bari}
_{ij}$ enter the formula for the radiated energy. Small deviations in
the waveforms therefore can account for quite large differences in the
energies.  It is not surprising, though not easy to explain in detail,
that the 3D waveforms differ from the axisymmetric waveforms at later
times $t \ga 33$\,ms, when the dynamical evolution of these models is
changed significantly due to their non-axisymmetric inner cores.

Given the dynamical evolution of the core, the small magnitude of the
amplitudes $h_\times$ can easily be explained by utilizing the well
known order-of-magnitude argument for the quadrupole waveforms

\beq
rh_\times\simeq 2\,\frac{G}{c^4}\frac{I \bari}{T^2}\simeq
2\,\frac{G}{c^4}\,\frac{0.1 M R^2}{T^2}\simeq 100\mbox{\,cm} \,.
\eeq

According to the results of our computations, we have inserted the
quadrupole moment $I \bari$ of a homogeneous bar with mass $M \sim
0.5\ms$, length $R \sim 100$\,km and rotation period $T \sim
5$\,ms.  Approximating the non-axisymmetrically distributed mass of
$\sim 0.5\ms$ as two point masses orbiting each other at a distance
of 100\,km with an angular velocity equal to the observed one, yields
the same order-of-magnitude for $h_\times$. 

Finally, by using the quadrupole formulae (Eqns.\,\ref{hplus} and
\ref{hcross}) we may have underestimated the amount of gravitational
radiation produced by model MD2, particularly during the phase when
the $m=3$ symmetry of the inner core is most prominent.  We estimate
the order-of-magnitude of the mass-octupole contribution to the signal
by (\eg Blan\-chet \etal 1990, Eq.\,6.8)

\beq
\label{oct}
rh^{\rm TT}_{\rm oct} \simeq  2\frac{G}{c^4}\frac{3}{c}\frac{MR^3}{T^3}
                      \simeq  3R_{\rm s}\left(\frac{v}{c}\right)^3
                      \la     500 \mbox{\,cm}  \,,
\eeq

where $v \sim 0.1\,c$ and $M \sim 0.5\ms$ and a Schwarzschild radius
$R_{\rm s} \sim 1.5\,10^{5}$\,cm was assumed.  Hence, the
mass-octupole radiation cannot account for significant enhancement of
the peak amplitudes of our model MD2 compared with the axisymmetric
case (Fig.\,\ref{fig:3Dh}) or MD1, although it might change the
details of the waveforms.

\begin{figure}[!t]
\epsfxsize=8.8cm \epsfclipon \epsffile{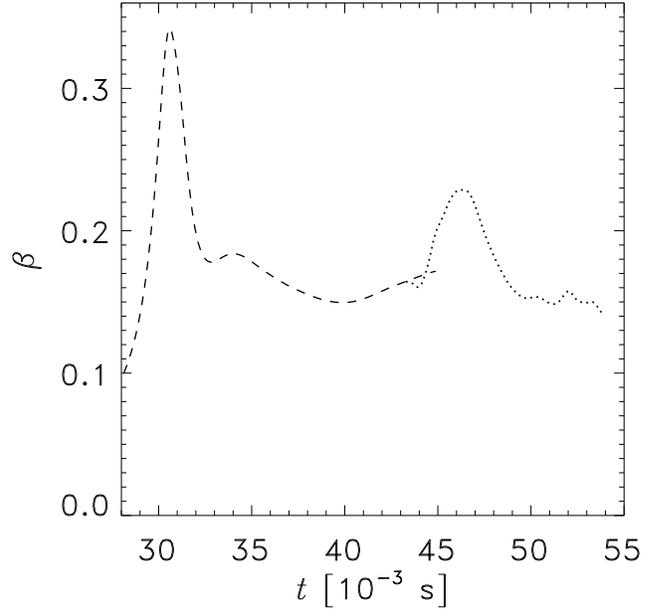}
\caption[]{Rotation parameter $\beta$ as a function of time for the
three-dimensional model MD3 (dotted), which suffers a second collapse.
For comparison the evolution of $\beta$ of model MD1 is shown, too
(dashed).}
\label{fig:MD3_beta}
\end{figure}

\subsection{Second collapse of a non-axisymmetric core}

After bounce, the proto-neutron star settles into its new rotational
equilibrium, and cools and contracts on a secular time scale, which is
given by the neutrino-loss time scale ($\simeq 10$\,s; see \eg Keil \&
Janka 1995 and references cited therein).  According to linear
stability analysis tri-axial perturbations grow on this time scale,
provided the rotation parameter $\beta \ga 0.14$.  For numerical
reasons (because our hydrodynamic code is explicit and thus the time
step is limited by the CFL stability condition; see \eg LeVeque 1992)
as well as for physical reasons (neglect of weak interactions and
neutrino transport) our approach is inadequate for simulating this
secular evolution.  However, we are able to consider a {\em sudden}
reduction of the stabilizing pressure in a deformed, non-axisymmetric,
rapidly rotating post-bounce core (model MD3).  This extreme case
cannot provide the answer to the question whether secular
instabilities will indeed grow to nonlinear amplitudes and what their
influence will be on the evolution of the core.  Nevertheless, such a
simulation can shed some light on the problem how much gravitational
radiation can be expected, when a rapidly rotating non-axisymmetric
neutron star forms.

\begin{figure*}[!t]
 \begin{tabular}{cc}
\put(0.9,0.3){{\LARGE\bf a}}
  \epsfxsize=8.8cm \epsfclipon \epsffile{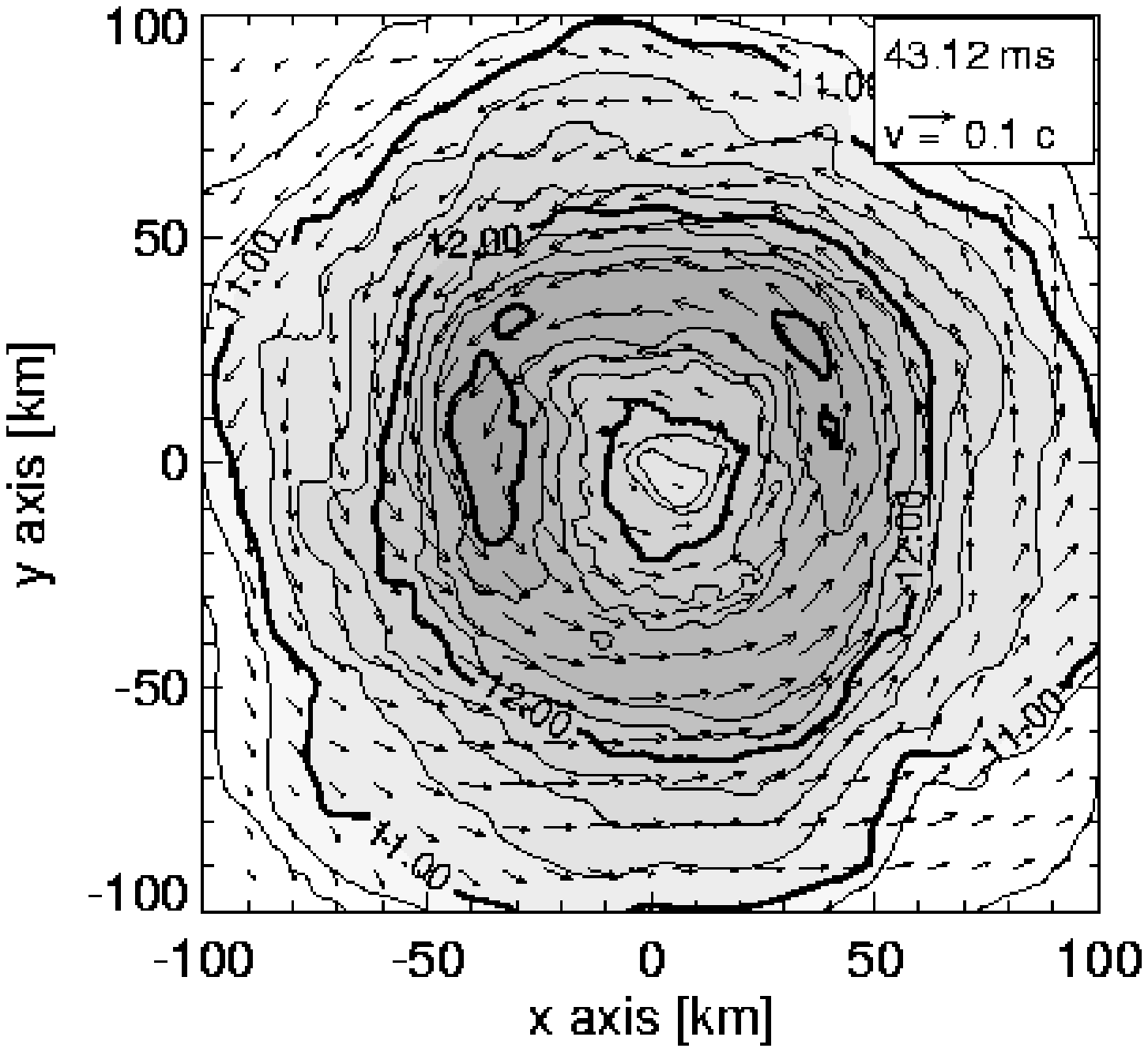} &
\put(0.9,0.3){{\LARGE\bf b}}
  \epsfxsize=8.8cm \epsfclipon \epsffile{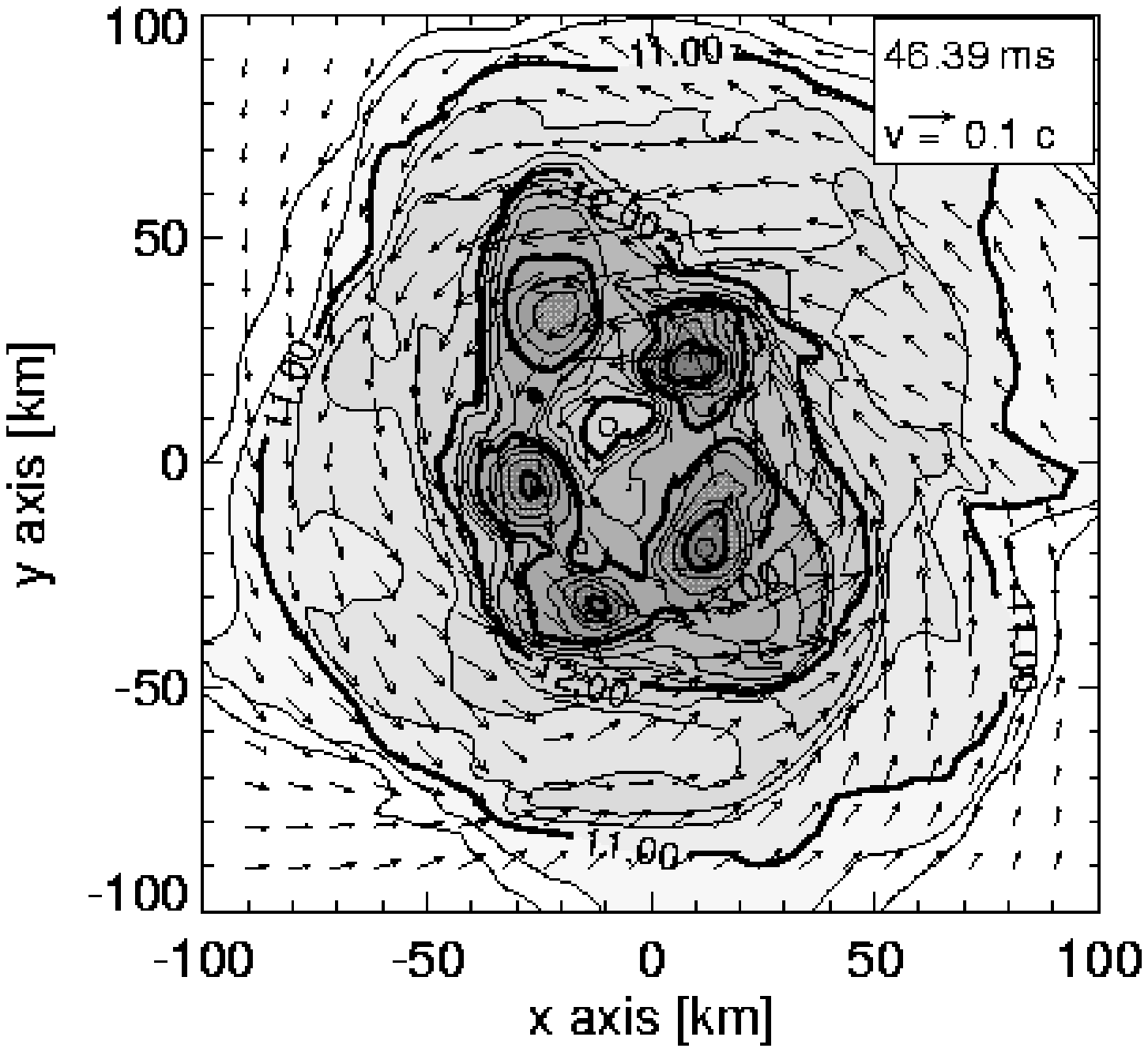} \\
\put(0.9,0.3){{\LARGE\bf c}}
  \epsfxsize=8.8cm \epsfclipon \epsffile{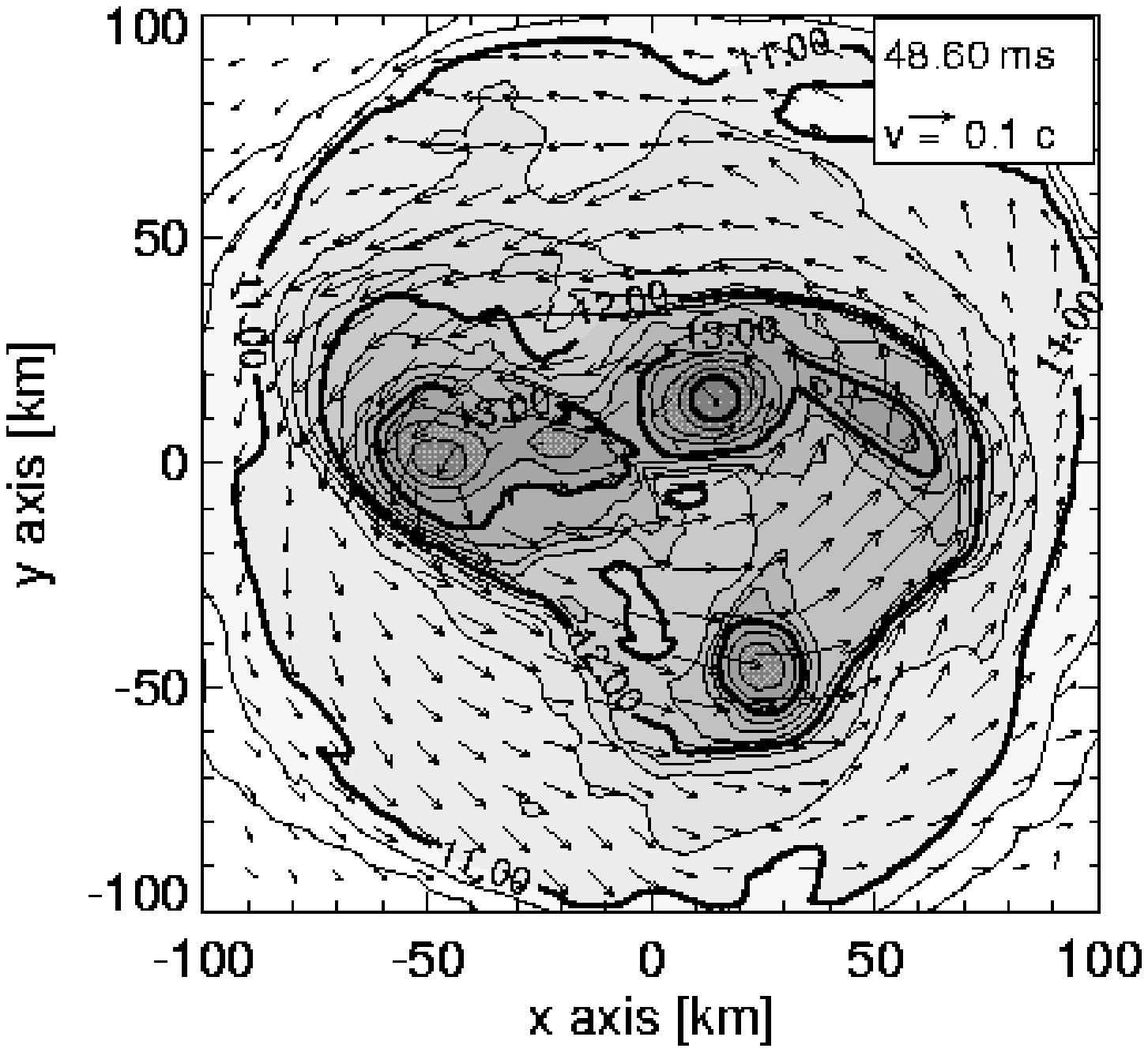} &
\put (0.9,0.3){{\LARGE\bf d}}
  \epsfxsize=8.8cm \epsfclipon \epsffile{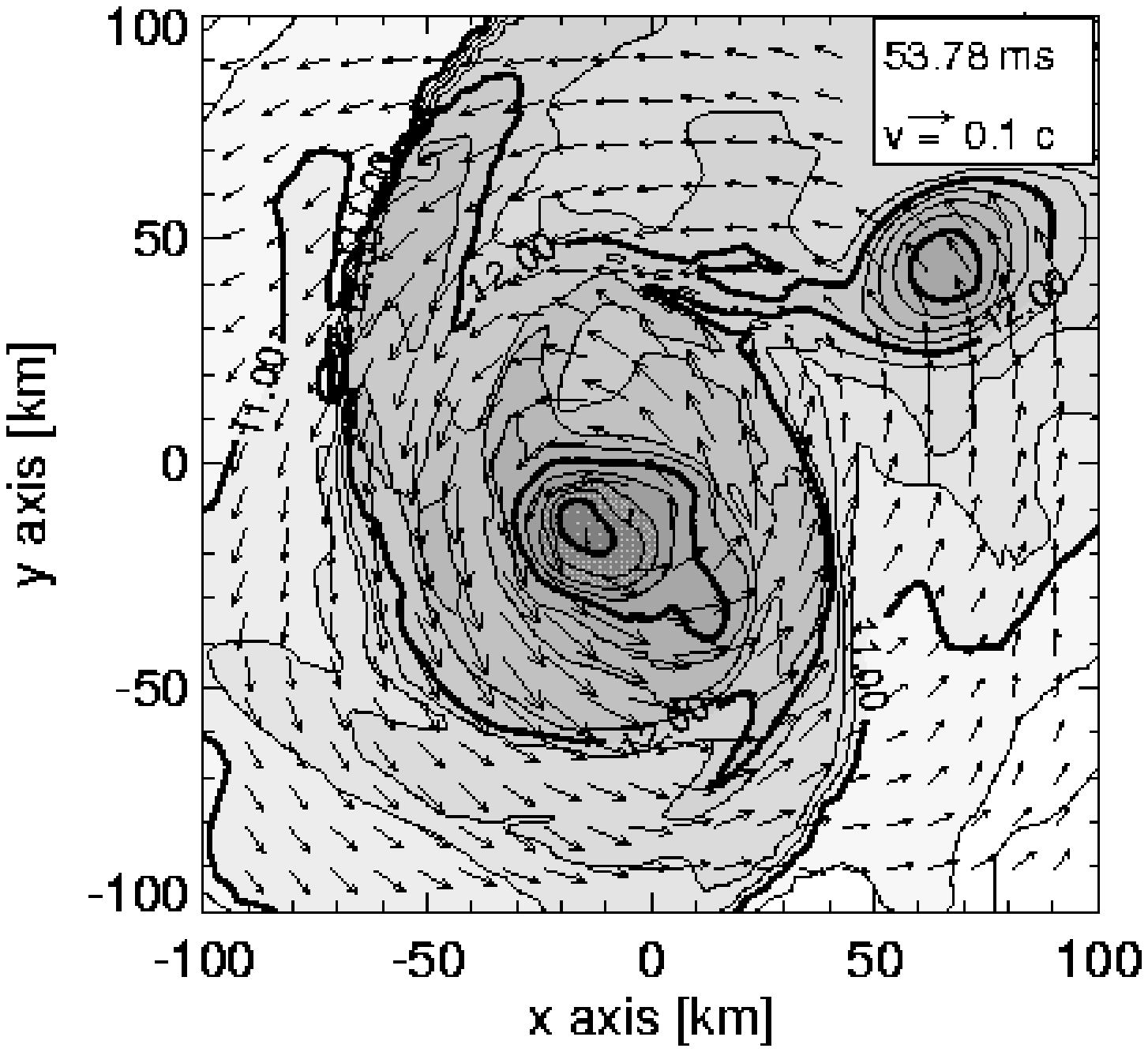} \\[-2ex]
 \end{tabular}
\caption[]{Snapshots of the density distribution (in units of
[$\gcc$]) and the velocity field in the equatorial plane of model
MD3. The contours are logarithmically spaced with intervals of 0.25
dex, they are shaded with darker grey values for higher density
regions and labeled with their respective values. The time of the
snapshot and the velocity scale are given in the upper right corner of
each panel.}
\label{fig:MD3_cont}
\end{figure*}

We choose model MD1 at $t=43.12$\,ms as the starting point for the
simulation (Fig.\,\ref{fig:MD3_cont}a).  At this time we (suddenly)
reduce $\Gamma_1$ from its original value of 1.28 to a value of 1.2.  
This
renders the rotating core unstable against a second dynamic collapse.
In passing we note that the stability criterion due to Ledoux (1945;
Eq.\,77) would require $\Gamma_1 \la 1.18$ for a rigidly rotating core
{\em in equilibrium} with $\beta \approx 0.16$ to collapse.

\begin{figure}[!h]
\epsfxsize=8.8cm \epsfclipon \epsffile{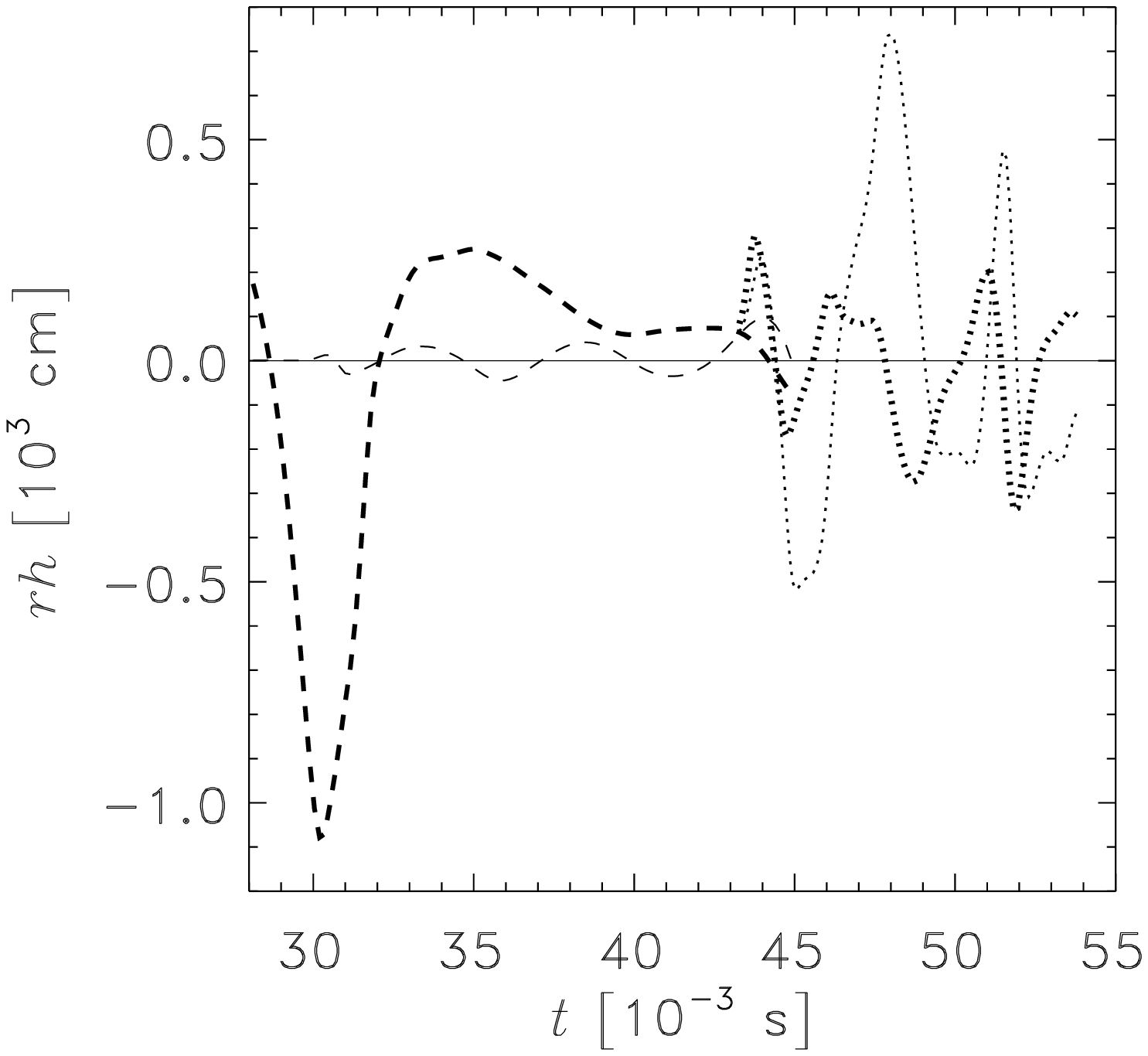}
\caption[]{Gravitational wave amplitudes $rh_+ (x\teq r, y\teq 0,
z\teq 0)$, (thick lines) and \mbox{$rh_{\times} (x\teq 0, y\teq 0,
z\teq r)$}, (thin lines) as functions of time for the
three-dimensional model MD3 (dotted), where $r$ denotes the distance
to the source. For comparison the amplitudes of model MD1 are shown,
too (dashed).}
\label{fig:2Ch}
\end{figure}

The overall contraction, bounce and re-expansion of model MD3 is
reflected in the time evolution of the rotation parameter $\beta$.
Note, that concerning the development of non-axisymmetric
instabilities the absolute value of $\beta$ is not very relevant here,
because the initial ``perturbations'' are already in the non-linear
regime.  Compared to the (first) collapse of model MD1 we find a
somewhat longer time scale for the contraction and re-expansion in
model MD3. The peak value of $\beta \approx 0.23$ is considerably
smaller than that of model MD1 ($\beta \approx 0.34$), because only
mass shells with $M \la 0.5\ms$ contract to radii similar to those
reached during the collapse of model MD1. Mass shells of model MD3
with $M \ga 0.8\ms$ remain nearly unaffected by the sudden pressure
reduction at $t=43.12$\,ms.

During the collapse of model MD3 several pronounced density maxima are
formed.  They are all located inside the torus reaching peak values of
$\varrho = 5.3\,10^{14}\gcc$.  The five initial clumps forming at the
density maxima contain a mass of $\approx 0.1\ms$ each
(Fig.\,\ref{fig:MD3_cont}b).  During the further evolution their
number decreases as they merge one after the other.  Eventually, just
two of them remain, which form a bar-like or binary-like deformed
central region (Fig.\,\ref{fig:MD3_cont}c and d).

The maximum density on the grid (reached in the clumps) sharply rises
from $2\,10^{13}\gcc$ to $5.3\,10^{14}\gcc$ within a fraction of a
millisecond after pressure reduction.  The collapse is, however, rapidly
decelerated by the action of centrifugal forces, which also eventually
cause a bounce at $t \approx 46$\,ms (Fig.\,\ref{fig:MD3_beta}).  The
pressure increase resulting from matter whose density exceeds nuclear
matter density is dynamically unimportant, as only a negligible amount
of mass inside some of the clumps is involved
(Fig.\,\ref{fig:MD3_cont}).

Model MD3 shows a highly non-axisymmetric, bar-like density
stratification already {\em before} the (second) bounce occurs at
$t\approx 46$\,ms (Fig.\,\ref{fig:MD3_cont}b).  This is different from
the behaviour of models MD1 and MD2, where large non-axisymmetries
(quadrupole moments) only appear well {\em after} bounce, \ie after the most
rapid phase of the evolution.  Accordingly, one expects larger time
derivatives of larger quadrupole moments and therefore a stronger
gravitational wave signal from model MD3 than either from model MD1 or
MD2.

The actual waveforms for model MD3 are shown in Fig.\,\ref{fig:2Ch}.
Compared to model MD1 one notices a considerably larger value for the
genuinely non-axisymmetric polarization amplitude $h_\times$.  The
maximum values $|rh_\times| \approx 800$\,cm and $|rh_+| \approx
300$\,cm do, however, neither exceed the peak value of $|rh_+|$
obtained in the axisymmetric simulation at bounce nor those of the
three-dimensional models MD1 and MD2 (Fig.\,\ref{fig:2Ch}).

Additional (axisymmetric) simulations with $\Gamma_1$ reduced to 1.2 or
1.25 have shown that the peak value of $\beta$, the time scale of the
(second) collapse as well as the peak values of the gravitational
waveforms do not depend  strongly on the exact value of $\Gamma_1$.

\section{Summary and discussion}

We have presented {\em three-dimensional} hydrodynamic simulations of
the collapse of rapidly rotating stellar iron cores. The matter in the
core has been described by a simple analytical equation of state. 
Effects due to neutrino transport have been neglected.  Hydrodynamics
and self-gravity has been treated in Newtonian approximation.  The
initial models for our simulations have been taken from the
comprehensive parameter study of {\em axisymmetric} rotational core
collapse performed by Zwerger \& M\"uller (1997). 

For our study we have selected those axisymmetric models which are
secularly or dynamically unstable with respect to non-axisymmetric
perturbations. Hence, in order to be selected the rotation rate
parameter $\beta$ of a model had to exceed the critical rotation rates
$\beta_{\rm sec} \approx 0.14$ or $\beta_{\rm dyn} \approx 0.27$
during its evolution. These critical rotation rates have been derived
for incompressible MacLaurin spheroids, but also hold approximately
for a wide range of compressible {\em equilibrium} configurations (\eg
Tassoul 1978).  After mapping the rapidly rotating axisymmetric cores
onto the three-dimensional computational grid, we imposed an initial
non-axisymmetric perturbation and simulated the further evolution with
a three-dimensional variant of the PROMETHEUS hydrodynamic code. The
simulations cover a time interval from a few milliseconds before core
bounce up to several tens of milliseconds after bounce.

We have used the quadrupole formula to calculate the gravitational
wave signal and have estimated the contribution of the next-leading
radiative-multipole order, which is negligibly small.  Because of its
smallness gravitational radiation reaction was not taken into account.

Our 3D simulations show that in two models, which are secularly but
not dynamically unstable ($0.15 \le \beta \le 0.2$) non-axisym\-metric
perturbations do not grow. Consistent with this result, we also
observe no significant enhancement of the gravitational wave emission
in these models compared to the axisymmetric case.

Among the models investigated by Zwerger \& M\"uller (1997) there is
only one model where $\beta > \beta_{\rm dyn}$ during the
evolution. This is their most rapidly and most differentially rotating
model evolved with the softest equation of state.  This axisymmetric
model was perturbed about 2\,ms before bounce by imposing a random
density distribution of 10\% amplitude.  Despite the random initial
perturbation, $m=2$ and $m=4$ toroidal modes are found to dominate the
early evolution of the model, because of the set of cubic grids used
in the simulation.  Therefore, a second simulation was performed,
where an additional $m=3$ toroidal perturbation of 5\% amplitude was
imposed initially.

The gross features of the evolution are quite similar in the
axisymmetric and non-axisymmetric simulations.  In both cases the
overall evolution is characterized by a rapid contraction, bounce and
re-expansion of the inner core.  In the 3D simulations we, in
addition, observe the growth of the initial perturbations.  Until
several milliseconds (or dynamical time scales of the inner core)
after bounce the density distribution resembles the symmetry of the
initially dominating modes. Subsequently, the core evolves towards a
bar-like shaped configuration. In none of the 3D models the
gravitational wave amplitude exceeds the axisymmetric value of
$|h|=3.5\,10^{-23}$ (for a source at 10\,Mpc), three dimensional
effects on the waveforms being only of the order of 10\%.  Even when
the collapsed, bar-like, rapidly rotating inner core is forced into a
second collapse (by artificially reducing its adiabatic index), we do
not observe gravitational wave amplitudes significantly larger than
$|h|\simeq 10^{-23}$ (for a source at 10\,Mpc).

In order to judge the implications of our results for rotational core
collapse in general and for gravitational wave astronomy in
particular, the following possible limitations of our approach should
be kept in mind:

\begin{itemize}

\item[(i)] Since $GM/Rc^2\la 0.2$ for the axisymmetric models of
Zwerger \& M\"uller (1997), general relativistic effects can be viewed
as moderate corrections to Newtonian gravity as far as the collapse
dynamics is concerned. However, since general relativity counteracts
the stabilizing influence of rotation on radial modes (\eg Tassoul
1978), its influence on the stability properties of (even only
moderately compact) rotating iron cores can be of considerable
importance. When GR is taken into account, pre-collapse models with a
larger amount of angular momentum (and the same EOS) than
in the Newtonian approximation can collapse. Furthermore, GR models will maintain
higher densities as well as larger rotation parameters for a longer
time interval after bounce as compared to Newtonian models (see the
discussion in Zwerger \& M\"uller 1997).

\item[(ii)] The {\em secular evolution} (\ie the evolution on time
scales of hundreds of milliseconds) of the collapsed core is not known
even in the axisymmetric case.  Simulating this evolution requires
models with detailed microphysics and thermodynamics, and an adequate
treatment of the neutrino transport. However, even in 2D such
simulations are still prohibitively time consuming.  Thus, there might
exist additional initial models different from those presently
available, which could become unstable to tri-axial perturbations on a
secular time scale (see however the next point).

\item[(iii)] Concerning the {\em limited set of axisymmetric initial
models}, we adopt the arguments given in Zwerger \& M\"uller
(1997). They claim that the large parameter space considered in their
study most probably comprises the whole domain where actual
pre-collapse rotating iron cores are to be found. They further argue
that using a realistic equation of state instead of a polytropic one
will change the details of the evolution, but will not give rise to
qualitatively different results. In particular, it is unlikely that
there exists a much larger set of pre-collapse initial models, which
fulfill $\beta > \beta_{\rm dyn}$ {\em and} which are sufficiently
compact for an extended time interval after core bounce.

But even if there are models which fulfill these conditions
and become triaxial, we do
not expect them to produce a considerably stronger gravitational wave
signal than the ``best'' axisymmetric
models. Provided the initial amount of angular momentum is not
unreasonably large, the critical rotation rates for non-axisymmetric
instabilities can only be exceeded in collapsing cores with relatively
soft equations of state (Eriguchi \& M\"uller 1985, Zwerger \&
M\"uller 1997).  But these soft equations of state drastically reduce
both the mass and the radius of the inner core (Zwerger \& M\"uller
1997).  However, according to our 3D calculations, it is only in the
inner core, where non-axisymmetric perturbations can grow
significantly.  Ta-ken together these considerations suggest that
bar-like inner cores have masses less than $0.5\ms$, radii less than
$100$\,km, and rotation periods greater than $1$\,ms. This implies a
maximum gravitational wave amplitude $|h| \la 10^{-22}$ (for a source
at 10\,Mpc).
\end{itemize}

Given the most recent rates of core collapse supernova events (\eg
Cappellaro \etal 1997), which not all may involve a rapidly rotating
core, we conclude that it is rather unlikely to expect gravitational
wave signals from dynamic non-axisymmetric instabilities with
sufficient strength {\em and} rate (\ie several per year) to be
observable with the large interferometric gravitational wave antennas
presently under construction.

Movies in MPEG format of the dynamical evolution of all models are
available in the world-wide-web at {\sl http:
//\newline www.mpa-garching.mpg.de/\lower0.7ex\hbox{$\!$\~~$\!$}mjr/GRAV/grav3.html}

\begin{acknowledgements}
The calculations were performed at the Rechenzentrum Garching on a
Cray J90. This research was supported in part by the National Science
Foundation under Grant No.~PHY~94-07194.
\end{acknowledgements}

\end{document}